\documentclass[twocolumn,aps,pre,amsmath,nofootinbib,longbibliography]{revtex4-1}
\usepackage{float}
\usepackage{color}
\usepackage{graphicx}
\usepackage{ulem}
\usepackage{epstopdf}
\usepackage{amsfonts}
\usepackage{amsmath}
\usepackage{mathtools}
\usepackage{bbold}
\usepackage[toc,page]{appendix}
\usepackage{enumerate}
\usepackage{comment}
\newcommand{\diag}{\mathop{\rm diag}}

\newcommand{\ec}{\overline{c}}
\newcommand{\eC}{\overline{\mathbf{C}}}
\newcommand{\eCr}{\overline{\mathbf{C}}_{[{\rm r},{\rm r}]}}
\newcommand{\nc}{c^{*}}
\newcommand{\nC}{\mathbf{C}^{*}}
\newcommand{\mean}[1]{\left\langle #1 \right\rangle}

\newcommand{\x}{\mathbf{x}}
\newcommand{\xr}{\mathbf{x}_{\rm r}}

\newcommand{\y}{\mathbf{y}}
\newcommand{\yr}{\mathbf{y}_{\rm r}}

\newcommand{\A}{\mathbf{A}}

\newcommand{\Aeff}{\mathbf{A}_\textup{eff}}

\newcommand{\C}{\mathbf{C}}
\newcommand{\Cr}{\mathbf{C}_{[{\rm r},{\rm r}]}}
\newcommand{\D}{\mathbf{D}}
\newcommand{\Dr}{\mathbf{D}_{[{\rm r},{\rm r}]}}
\newcommand{\F}{\mathbf{f}}
\newcommand{\Fr}{\mathbf{f}_{\rm r}}
\newcommand{\I}{\mathbf{I}}
\newcommand{\J}{\mathbf{j}}

\newcommand{\Om}{\mathbf{\Omega}}
\newcommand{\Omr}{\mathbf{\Omega}_{\rm r}}
\newcommand{\V}{\mathbf{v}}
\newcommand{\Vr}{\mathbf{v}_{\rm r}}

\newcommand{\pdx}{\widetilde{\partial}_{1}^2}
\newcommand{\pdy}{\widetilde{\partial}_{2}^2}

\newcommand{\nb}{\bar{n}}
\newcommand{\N}{\mathbf{n}}
\newcommand{\Nb}{\bar{\mathbf{n}}}
\newcommand{\tz}{\tilde{z}}
\newcommand{\sd}{\sum_{i=1}^{d}}

\newcommand{\pdni}{\tilde{\partial}_{n_i}^2}
\newcommand{\pdnib}{\tilde{\partial}_{\bar{n}_i}^2}

\newcommand{\parib}{\tilde{\partial}_{\bar{1}}^2}
\newcommand{\parjb}{\tilde{\partial}_{\bar{2}}^2}
\newcommand{\pardb}{\tilde{\partial}_{\bar{3}}^2}

\newcommand{\pari}{\tilde{\partial}_{1}^2}
\newcommand{\parj}{\tilde{\partial}_{2}^2}
\newcommand{\pard}{\tilde{\partial}_{3}^2}

\newcommand\chout{\bgroup\markoverwith{\textcolor{red}{\rule[0.5ex]{2pt}{1.0pt}}}\ULon}
\newcommand\fedout{\bgroup\markoverwith{\textcolor{blue}{\rule[0.5ex]{2pt}{1.0pt}}}\ULon}
\newcommand\grzout{\bgroup\markoverwith{\textcolor{green}{\rule[0.5ex]{2pt}{1.0pt}}}\ULon}

\begin{document}

\title{Scaling behavior of non-equilibrium measures in internally driven elastic assemblies}

\author{Grzegorz Gradziuk}
\affiliation{Arnold-Sommerfeld-Center for Theoretical Physics and Center for
  NanoScience, Ludwig-Maximilians-Universit\"at M\"unchen,
   D-80333 M\"unchen, Germany.}
   \author{Federica Mura}
\affiliation{Arnold-Sommerfeld-Center for Theoretical Physics and Center for
  NanoScience, Ludwig-Maximilians-Universit\"at M\"unchen,
   D-80333 M\"unchen, Germany.}
\author{Chase P. Broedersz}
\email{C.broedersz@lmu.de}
\affiliation{Arnold-Sommerfeld-Center for Theoretical Physics and Center for
  NanoScience, Ludwig-Maximilians-Universit\"at M\"unchen,
   D-80333 M\"unchen, Germany.}

\date{\today}

\begin{abstract}
Detecting and quantifying non-equilibrium activity is essential for studying internally driven assemblies, including synthetic active matter and complex living systems such as cells or tissue. We discuss a non-invasive approach of measuring non-equilibrium behavior based on the breaking of detailed balance. We focus on ``cycling frequencies" - the average frequency with which the trajectories of pairs of degrees of freedom revolve in phase space, and explain their connection with other non-equilibrium measures, including the area enclosing rate and the entropy production rate. We test our approach on simple toy-models comprised of elastic networks immersed in a viscous fluid with site-dependent internal driving. We prove both numerically and analytically that the cycling frequencies obey a power-law as a function of distance between the tracked degrees of freedom. Importantly, the behavior of the cycling frequencies contains information about the dimensionality of the system and the amplitude of active noise. The mapping we use in our analytical approach thus offers a convenient framework for predicting the behavior of two-point non-equilibrium measures for a given activity distribution in the network. 
\end{abstract}
\maketitle

\section{Introduction}
\label{sec:intro}
\noindent 

The field of active matter has developed over the last decades to provide a physical description of classical many-body systems operating far from thermodynamic equilibrium~\cite{Fodorreview,Jülicherreview,Marchettireview}. A prominent class of such active matter are living systems: Schools of fishes~\cite{Filella2018}, flocks of birds~\cite{Ballerini2008}, and colonies of bacteria~\cite{Damton2010,Thutupalli2015a} can all exhibit collective dynamics that are manifestly out of equilibrium. However, the non-equilibrium activity of biological assemblies at smaller subcelluar scales is not always straightforward to discern~\cite{mackintosh2010,Gnesottoreview}. Examples include the stochastic fluctuations of biological assemblies such as chromosomes~\cite{Weber}, the cytoskeleton~\cite{Lau2003,Guo2014,Fakhri2014,Jülicherreview,Brangwynne2008}, and cellular membranes~\cite{Betz2009,Turlier2016,Ben-Isaac2011}. Indeed, while these fluctuations can at first sight appear indistinguishable from thermal Brownian motion, they are in many cases driven by energy-consuming processes at molecular scales~\cite{Gnesottoreview,Fodorreview,Needleman2017b, Suzuki2015,mackintosh2010,Jülicherreview}.
This molecular scale activity can propagate to mesoscopic scales, giving rise to non-equilibrium dynamics that breaks detailed balance~\cite{Battle2016,Gladrow2016,Gladrow2017,Mura2018,
Gnesottoreview,Gnesotto2018,Li2018} or that violates the fluctuation dissipation theorem~\cite{Turlier2016,Betz2009,Mizuno2007,Martin2001,Fodor2015,Guo2014}. Soft driven assemblies can also be realized in synthetic systems, including chemical fueled synthetic fibers~\cite{Boekhoven2015} and crystals of active colloidal particles~\cite{Palacci2013}.
Numerous experimental studies showed how molecular non-equilibrium processes affect the mesoscopic mechanical properties of \textit{in vivo} biological assemblies~\cite{Lau2003,Guo2014,Fakhri2014,Jülicherreview},  \textit{in vitro} reconstituted cytoskeletal networks~\cite{Liu2006,Koenderink2009,Alvarado2017} and synthetic materials~\cite{Bertrand2012}.  
 It still remains unclear, however, how to characterize the non-equilibrium fluctuations of soft driven assemblies.

 To make further progress on characterizing active systems, various candidates for a reliable and informative non-equilibrium measure have been proposed. A natural and commonly used measure of the time-irreversibility of a process is the entropy production rate. In some cases, this measure is related to  the energy dissipation in a system~\cite{Seifertreview}. Recent studies made significant progress in inferring the entropy production rate from the observed trajectories~\cite{Li2018,Frishman2018,Seara2018}. In general, for complex systems it is unclear how to interpret measures of the partial entropy production rate or how to relate the measured quantities to the real entropy production rate of the full system. It is possible, however, to set a lower bound to the total entropy production from the observation of a few mesoscopic degrees of freedom~\cite{Roldan2018,Mura2018, Bisker2017,Esposito2012,Polettini2017}. An alternative approach of using area enclosing rates (AER) of stochastic trajectories in phase space as a metric for the breaking of detailed balance was presented in~\cite{Ghanta2017} and~\cite{Gonzalez2018}. A closely related concept---the cycling frequencies of the stochastic trajectory---was used to analyse the non-equilibrium behavior emerging in the mode-space trajectories of a probe filament in an active gel~\cite{Gladrow2016,Gladrow2017} or the dynamics of  a driven disordered elastic 
network near its isostatic point~\cite{Gnesotto2018}.

Despite the multitude of available non-equilibrium measures, it is still unclear how to use them to extract useful information about the nature of active driving in a system. The cycling frequencies can be used to investigate the  non-equilibrium dynamics emerging on different lengthscales in driven elastic networks~\cite{Mura2018}. In particular, the cycling frequencies measured from trajectories of two probe particles in an internally driven elastic network display a power-law behavior as function of the distance between the particles, with an exponent that depends on the dimensionality of the system; the prefactor of this scaling law depends on the statistical properties of the internal driving. Thus these experimentally accessible cycling frequencies and their associated scaling behavior provide a promising candidate for a non-equilibrium measure that may give provide access to properties of the internal driving.

In this work we present a detailed derivation of the scaling behavior of cycling frequencies for $d$-dimensional elastic networks with internal driving. Thus, we derive a theoretical framework that allows us to relate the cycling frequency to the lengthscale of the observation and to the properties of the network and of the active noise. Furthermore, we clarify the relation between the cycling frequencies and other non-equilibrium measures such as the area enclosing rate and the entropy production rate. 
\begin{figure}
\centering
\includegraphics[width=8cm]{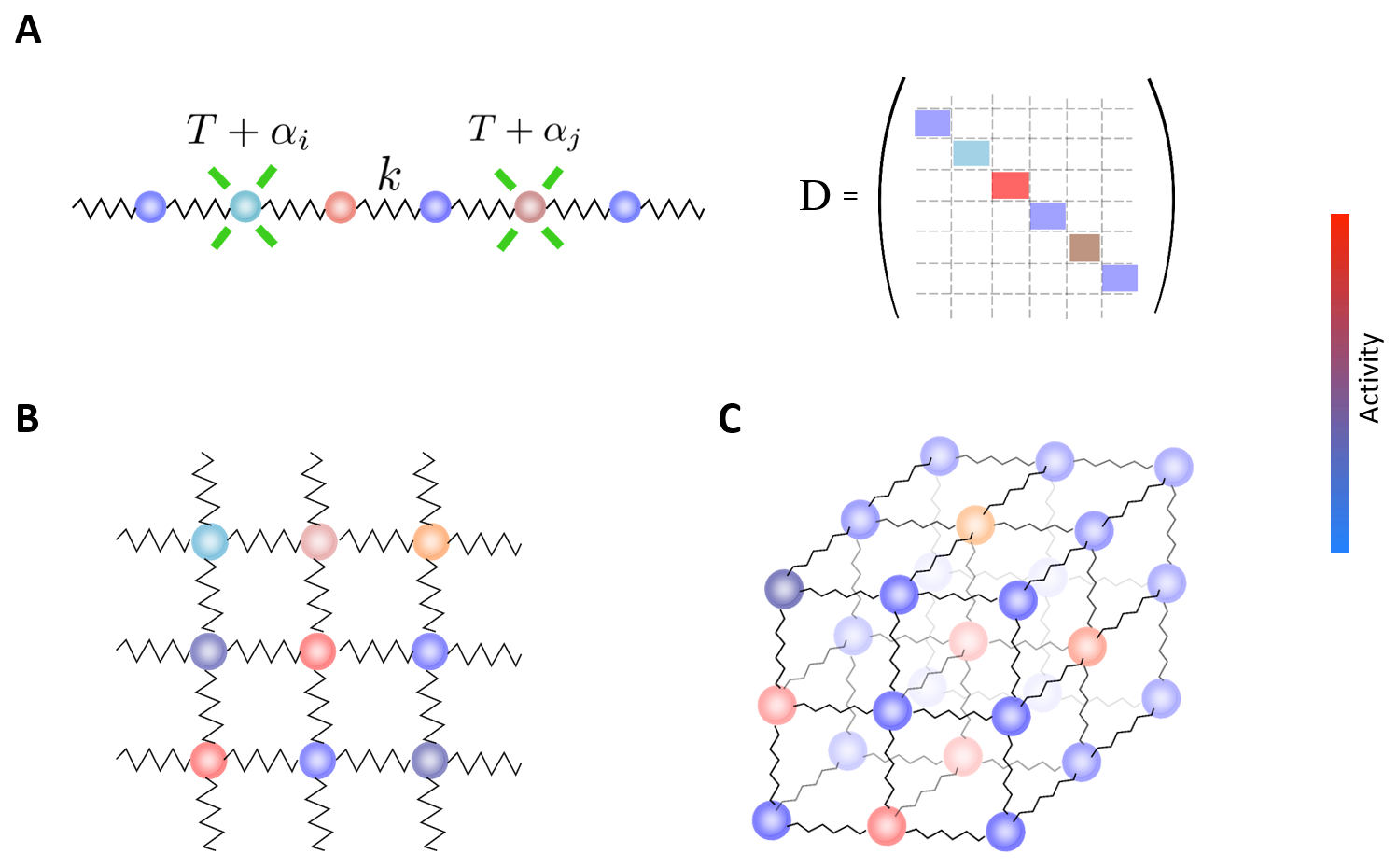}
\caption{
A) 1-dimensional elastic chain of beads at temperature $T$ with spatially varying white-noise driving intensity $\alpha_i$, and the corresponding diffusion matrix $D$. The tracked pair of beads is indicated in green. B) 2-dimensional and C) 3-dimensional elastic networks investigated in Sec.~\ref{sec:Cubic lattice}.}
\label{fig:intropanel}
\end{figure}
\section{Driven elastic networks}
\label{sec1}
We use overdamped networks of elastically coupled beads suspended in a viscous fluid as a simple model for soft subcellular  assemblies~\cite{Yucht2013,Broedersz2014,Osmanovic2017,Mao2018}. For simplicity we choose units in which the elastic spring constant, the damping coefficient of the beads and the Boltzmann constant $k_B$ equal 1. The fluid is assumed to be at thermal equilibrium and the resulting thermal fluctuations in the system are thus modelled as Gaussian white noise processes acting independently on all the beads with the same amplitude $T$. Additional active force fluctuations drive the system out of thermal equilibrium, and are implemented as independent Gaussian white noise processes with spatially varying amplitudes $\alpha_i$. 

By modelling the active forces as ``white", we essentially restrict our model to systems in which the correlation times of the active driving are shorter than the intrinsic relaxation times of the network. This model is mathematically equivalent to embedding the beads in local thermal baths at temperatures $T+\alpha_i$~\cite{Falasco2015}.

This simplified description allows us to study the dynamics of the system using a Fokker-Planck equation:
\begin{equation}
\partial_t p(\x,t)=-\nabla \cdot (\A\x-\D\nabla)p(\x,t)\coloneqq-\nabla \cdot \J(\x,t), 
\label{eq:fokker}
\end{equation}
Here $\x$ represents the displacements of the beads relative to their equilibrium positions, $p(\x,t)$ is the probability distribution of $\x$ at time $t$. We also assume that the forces are linear in $\x$, i.e. $\F(\x)=\A\x$ with a symmetric matrix $\A$, and $\D=\diag\{T+\alpha_1,\ldots, T+\alpha_N\}$ is the diffusion matrix. The right hand side of the Eq.~\eqref{eq:fokker} can be interpreted as the divergence of the probability current density $\J(\x,t)=(\A\x-\D\nabla)p(\x,t)$. At  steady-state, the non-vanishing dissipative probability currents constitute a measure for non-equilibrium in a system and thus play the key role in our approach.

\section{Cycling frequencies \& phase space torque}
\label{sec2}
The steady-state probability currents, $\J(\x)$, are mathematical objects that capture the presence of non-equilibrium activity by revealing time-irreversibility of the dynamics at the level of the Fokker-Planck equation. This time-irreversibility manifests through the emergence of a mean velocity field $\V(\x)$ in the coordinate space, which is related to the probability current through $\J(\x)=\V(\x)p(\x)$~\cite{Risken}. Therefore, from an experimental perspective, an ideal way to quantify the non-equilibrium dynamics of a system would be to measure such a velocity field $\V(\x)$. However, inferring the full $\V(\x)$ field is a challenge on its own. The most straightforward approaches require a discretisation of the phase space. Such a measurement would require tracking many degrees of freedom for long time periods, which is difficult in practice.

\begin{figure}
\centering
\includegraphics[width=0.8\linewidth]{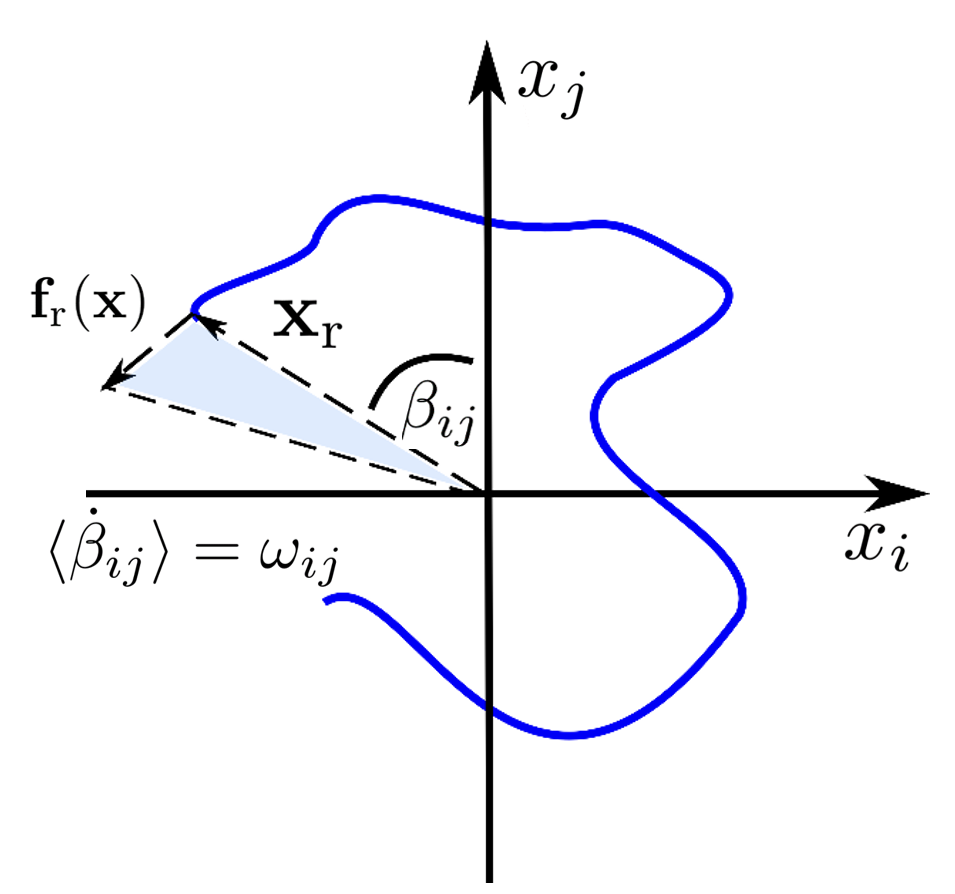}
\caption{ Schematic trajectory in the coordinate space $[x_i,x_j]$ of two tracer beads. The light blue area enclosed in the triangle represents $\xr\times\Fr(\x)/2$ appearing in Eq.~\eqref{eq:torque}, which gives the area enclosing rate upon averaging over phase space.}
\label{fig:torque}
\end{figure}

Instead of inferring $\V(\x)$ in full detail, one can alternatively measure some coarse-grained quantities related to $\V(\x)$, which still retain key information about the non-equilibrium dynamics of the system. For instance, we could track a pair of degrees of freedom $\xr=\{x_i,x_j\}$ and measure the average angular velocity $\langle \dot{\beta}_{ij}\rangle$, or equivalently, the rate at which the trajectory revolves around the origin in this reduced 2-dimensional subspace (Fig.~\ref{fig:torque}). This simple measurement does not require any discretisation of phase space or inference of the force field. We shall refer to $\langle \dot{\beta}_{ij}\rangle$ as the \textit{cycling frequency}. 

In general, $\langle \dot{\beta}_{ij}\rangle$ may contain only limited information about $\V(\x)$. For linear systems, however, the mean phase space velocity can be written as~\cite{Weiss2003}
\begin{equation}
\V(\x)=\Om\x,\quad \Om=\A+\D\C^{-1}
\label{eq:omega}
\end{equation}
where $\C=\langle \x\otimes\x\rangle$ is the steady-state covariance matrix obeying the Lyapunov equation
\begin{equation}
\A\C+\C\A^T=-2\D.
\label{eq:Lyapunov}
\end{equation}
Equation~\eqref{eq:omega} sets strong constraints on the structure of $\V(\x)$: for dynamics projected on any 2-dimensional subspace $\{x_i,x_j\}$ the probability currents have an elliptical structure. The remaining information about the amplitude of the currents is set by the cycling frequencies $\langle \dot{\beta}_{ij} \rangle$.

To show this, we denote the velocity of the system in the reduced $ij$-subspace by $\Vr(\x)=\{ v_i(\x),v_j(\x)\}$. Note, what we observe while looking at the $\x_r$-subspace only is a conditional mean $\langle \Vr(\x)|\x_r \rangle$. Similarly to Eq.~\eqref{eq:omega}, we find that $\langle\Vr(\x)|\xr\rangle=\Omr\xr$, with $\Omr=\Aeff+\Dr\Cr^{-1}$~\cite{Mura2018}. Here $\Aeff$ is a matrix such that $\langle \Fr(\x)|\xr\rangle=\langle \{f_i(\x),f_j(\x)\}|\xr\rangle=\Aeff\xr$; note, $\Cr$ and $\Dr$ are matrices of size $[2 \times 2]$, given by $\Cr=\{\{c_{ii},c_{ij}\},\{c_{ji},c_{jj}\}\}$ and $\Dr=\{\{d_{ii},d_{ij}\},\{d_{ji},d_{jj}\}\}.$

Next, we show that the eigenvalues of $\Omr$  coincide with the cycling frequencies $\langle \dot{\beta}_{ij} \rangle$. 
First, note that $\langle \dot{\beta}_{ij} \rangle$ is invariant under orientation preserving linear transformations of the reduced subspace. We can therefore work in \textit{covariance identity coordinates}, $\yr$, such that $\Cr=\I$. In this basis, $\Omr$ takes a particularly simple form~\cite{Weiss2003}
\begin{equation}
\Omr=\left(\begin{array}{cc}
0 & \omega_{ij} \\
-\omega_{ij} & 0 \\
\end{array}\right),
\label{eq:omegaCIC}
\end{equation}
with the imaginary parts of its eigenvalues on the antidiagonal. This form of $\Omr$ implies that for $\Cr=\I$, the probability current field has a circular structure. Using Eq.~\eqref{eq:omegaCIC}, we find that
\begin{align}
\label{eq:indepomeg}
\langle\dot{\beta}_{ij}|\y_r\rangle&=\mean{\left.\frac{\yr\times\dot{\y}_{\rm r}}{|\yr|^2}\right\vert\yr}=\frac{y_2\mean{\dot{y}_1|\y_r}-y_1\mean{\dot{y}_2|\y_r}}{y_1^2+y_2^2}\nonumber\\
&=\frac{y_2\mean{v_1(\y)|\y_r}-y_1\mean{v_2(\y)|\y_r}}{y_1^2+y_2^2}=\omega_{ij}.
\end{align}
This means that the conditional average of the angular velocity $\langle\dot{\beta}_{ij}|\y_r\rangle$ is $\y_r$-independent and equals $\omega_{ij}$ at all points in the reduced phase space. Hence averaging over $\yr$ leads to $\langle\dot{\beta}_{ij}\rangle=\omega_{ij}$, with $\omega_{ij}^2=\det\Omr$. 

Recently, new approaches have been developed to infer current fields in nonlinear systems by considering an expansion of the inferred force field~\cite{Frishman2018}. Up to first order these methods reduce to calculating area enclosing rates, which are indeed closely related to the cycling frequencies, as we discuss further below.

One of our central objectives is to derive a relation between the observed currents and the properties of the system and the active driving. 
Given that the cycling frequencies are set by $\omega_{ij}$---the imaginary parts of the eigenvalues of $\Omr$---we can make further progress by showing that for a general linear system~\cite{Mura2018}
\begin{equation}
\omega_{ij}=\frac{1}{2}\frac{\mean{\xr\times\Fr(\x)}}{\sqrt{\det\C_r}},
\label{eq:torque}
\end{equation}
where $\mean{\xr\times\Fr(\x)}=\langle x_if_j(\x)-x_jf_i(\x)\rangle=(\C\A^T-\A\C)_{ij}$ is the mean phase space torque (Fig.~\ref{fig:torque}). Intuitively, Eq.~\eqref{eq:torque} implies that for an overdamped linear system the mean phase space angular velocity is proportional to the mean phase space torque. A detailed derivation is presented in ~\cite{Mura2018}. Moreover, in covariance identity coordinates Eq.~\eqref{eq:torque} reduces to $\omega_{ij}=\frac{1}{2}(A^T-A)_{ij}=\Omega_{ij}$, in accord with previous studies~\cite{Weiss2003, Gladrow2016,Gladrow2017}. One can equivalently identify $(\C\A^T-\A\C)_{ij}$ with mean area enclosing rates in the $ij$-subspace, as considered in~\cite{Ghanta2017}. Here, we focus on the cycling frequencies, since they are more directly related to the probability currents, which constitute the basis of our work. In some instances, however, the area enclosing rates turn out to be particularly advantageous to work with. In these cases we  shall briefly discuss how switching to the area enclosing rates simplifies the analysis (See Sec.~\ref{sec3}).

Since the cycling frequencies contain information about the amplitudes of the phase space probability currents, they can be related to the entropy production rate. For linear systems, in covariance identity coordinates, the full entropy production rate can be expressed as a weighted sum of the cycling frequencies squared: $\Pi=\sum_{i-\mathrm{odd}}^n \omega_{i,i+1}^2[(\D^{-1})_{i,i}+(\D^{-1})_{i+1,i+1}]$. However, in many experimental contexts it is typically impossible to track all the degrees of freedom or to resolve all steps of a process. This practical limitation motivated the introduction of various measures of reduced/apparent entropy production rate~\cite{Roldan2018,Mura2018, Bisker2017,Esposito2012,Polettini2017}. For the case of a two-point measurement, as the ones discussed in this paper, one can consider a reduced entropy production rate directly related to the cycling frequency: $\Pi_{\rm r}^{(2)}~=~\omega_{ij}^2\mathrm{Tr} (\Cr\Dr^{-1})$, which gives a lower bound on the full contribution to the total entropy production rate from the observed pair of degrees of freedom~\cite{Mura2018}.

\section{One-dimensional chain and diffusion equation}
\label{sec3}
While Eq.~\eqref{eq:torque} sets a relation between the cycling frequencies observable in an experiment and the properties of both the network and the active noise distribution, it is not straightforward to explicitly derive these  properties from the cycling frequencies. In the following sections, we use Eq.~\eqref{eq:torque} to build a framework for extracting specific information about the system from the behavior of the cycling frequencies $\omega_{ij}$.

We first consider the simplest case of a 1-dimensional chain of $2N-1$ beads coupled by harmonic springs. This example will help us build intuition for more complex lattices. To obtain insight into how non-equilibrium behavior manifests at different length scales, we consider a two-point non-equilibrium measure. Specifically, we study how the cycling frequency in the subspace of displacements of two chosen beads $\{ x_i,x_{i+r} \}$ depends on the distance $r$ between these beads. We do this for two scenarios:
\begin{enumerate}[i)]
\item Active noise present only at single site
\item Random spatial distribution of activities $\{\alpha_i\}$
\end{enumerate}
In case i) we plot  the cycling frequency, $\omega_{\rm single}(r)$, between the active bead and another bead at distance $r$. In case ii) we consider an ensemble of active noise distributions $\{\alpha_i\}$, in which the amplitude of the active noise at each site is drawn randomly from a probability distribution $p_\alpha$, with mean $\bar{\alpha}$ and variance $\sigma_\alpha^2$. The amplitudes $\alpha_i$ are spatially uncorrelated. We then calculate $\langle \omega^2(r) \rangle$ - the squared cycling frequency between two beads at distance $r$ averaged over the activity distributions $\alpha_i$. 
\begin{figure}
\centering
\includegraphics[width=8cm]{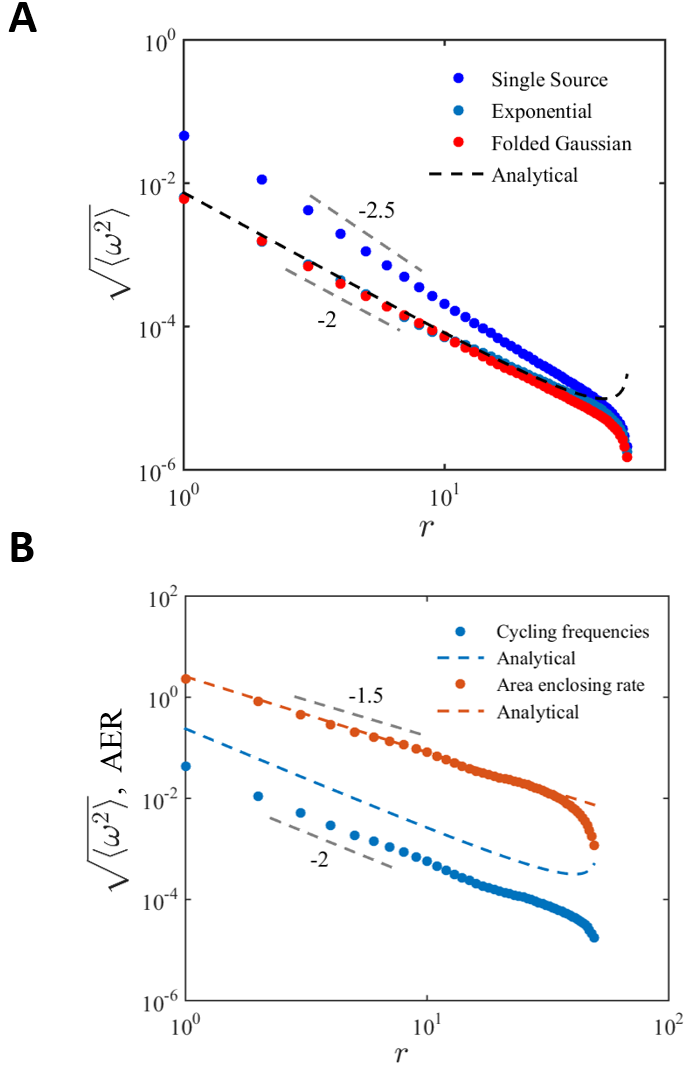}
\caption{A) Scaling behavior of the cycling frequencies as a fuction of distance between the beads, obtained for a 1D chain and different types of activity distributions, as indicated in the legend. B) Comparison between the cycling frequencies and the Area Enclosing Rates (AER) above the weak noise limit ($\bar{\alpha}/T=4$). All data points correspond to results obtained by numerically solving Lyapunov equation.}
\label{fig:omega1D}
\end{figure}
The first observation is that in both scenarios the cycling frequencies follow a power-law as a function of distance, as shown in Fig.~\ref{fig:omega1D}. The exponent in the random distribution scenario is independent of the probability distribution $p_\alpha$ of the intensities, but different from the single activity case. 

To understand the origin of the power-law behavior and to calculate the exponents, we use Eq.~\eqref{eq:torque} to derive analytical expressions for $\omega_{\rm single}(r)$ and $\langle\omega(r)^2\rangle$.
In the case of a 1-dimensional chain with spatially uncorrelated noise, the expression for the cycling frequency  (Eq.~\eqref{eq:torque}) reduces to:
\begin{equation}
\omega_{ij}=\frac{\pdy c_{ij}}{\sqrt{\det\Cr}},
\label{eq:omchain}
\end{equation}
where $c_{ij}$ indicates the elements of the covariance matrix, and $\pdy c_{ij}$ denotes the discrete second derivative across rows: $\pdy c_{ij}=c_{i,j+1}-2c_{i,j}+c_{i,j-1}$. Thus, this result reduces the problem of calculating $\omega_{ij}$ to finding the covariance matrix $\C$.

Motivated by the structure of $\D$, we decompose $\C=T\eC+\bar{\alpha}\nC$ into equilibrium $(\eC)$ and non-equilibrium $(\nC)$ parts (Fig.\ref{fig:numerC}A). For the 1-dimensional chain, the Lyapunov equation (see Eq.~\eqref{eq:Lyapunov}) is equivalent to:
\begin{align}
\pdx \ec_{ij}+\pdy \ec_{ij}&=-2\delta_{ij} \label{eq:diffec} \\
\pdx \nc_{ij}+\pdy \nc_{ij}&=-2\delta_{ij}\frac{\alpha_i}{\bar{\alpha}} \label{eq:diffnc},
\end{align}

At equilibrium detailed balance is preserved, which implies $\pdy \ec_{ij}=0\ \forall_{i\neq j}$ (see Eq.~\eqref{eq:omchain}). We can therefore replace $\pdy c_{ij}$ with $\pdy \nc_{ij}$ in Eq.~\eqref{eq:omchain}. Then, expanding the expression for the cycling frequency in powers of $(\bar{\alpha}/T)$, we get:
\begin{equation}
\omega_{ij}=\frac{\bar{\alpha}}{T}\frac{\pdy\nc_{ij}}{\sqrt{\det\eCr}}+\mathcal{O}\left(\frac{\bar{\alpha}^2}{T^2}\right)
\label{eq:omegalin}
\end{equation}
Up to linear order in $(\bar{\alpha}/T)$ the contributions from $\eC$ and $\nC$ separate; consequently $\omega_{ij}$ becomes linear in $\{\alpha_i\}$. This linearity appearing in the limit of weak activity will later allow us to calculate cycling frequencies averaged over different realizations of the activity $\{\alpha_i\}$. Note, if instead of $\omega_{ij}$, we consider the area enclosing rates ($\mathrm{AER}=\bar{\alpha}\pdy\nc_{ij}$), the factor $\sqrt{\det \Cr}$ does not enter, implying that the expression for the area enclosing rates is linear in $\{\alpha_i\}$ irrespective of the magnitude of $\bar{\alpha}$.

\subsection{Activity at a single site}
To obtain insight into what determines the cycling frequencies in a concrete example, we first find the solution to Eq.~\eqref{eq:diffnc} for the case of activity appearing only at a single site in the center of the chain. Later, we will use this solution to construct $\nC$ for a more general case. For now, let us assume that $\alpha_i=\alpha\delta_{iN}$ and $\bar{\alpha}=\alpha$. 

We can think of Eq.~\eqref{eq:diffnc} as a discretized stationary diffusion equation with a single source with a divergence of 2 in the middle of the $\nC$ matrix, and with absorbing boundary conditions at the edges. The absorbing boundaries in the diffusion equation reflect the fixed boundary conditions for the elastic chain. We denote by $r=\sqrt{(i-N)^2+(j-N)^2}$ the distance from the center of $\nC$. If we consider a continuous analogue of our discrete problem and neglect the boundary conditions, we can assume a rotational symmetry of the solution $\nc(r)$. The corresponding continuous diffusion equation takes the form: $\frac{1}{r}\partial_r r\partial_r \nc(r)=0$. Consequently, $r\partial_r\nc(r)=-a$ and $\nc(r)=-a\ln(r)+b$. 

One could also argue for $\partial_r\nc(r)\sim 1/r$ scaling of the ``covariance current", by demanding that the total ``covariance flux" through a circle of radius $r$, centred at $0$ is independent of $r$ and equals 2---the divergence of the source. This also allows us to identify $a=1/\pi$. The integration constant $b$ is system-size dependent and has to be set such that the covariance vanishes at the edge of the covariance matrix.

\begin{figure}
\centering
\includegraphics[width=8cm]{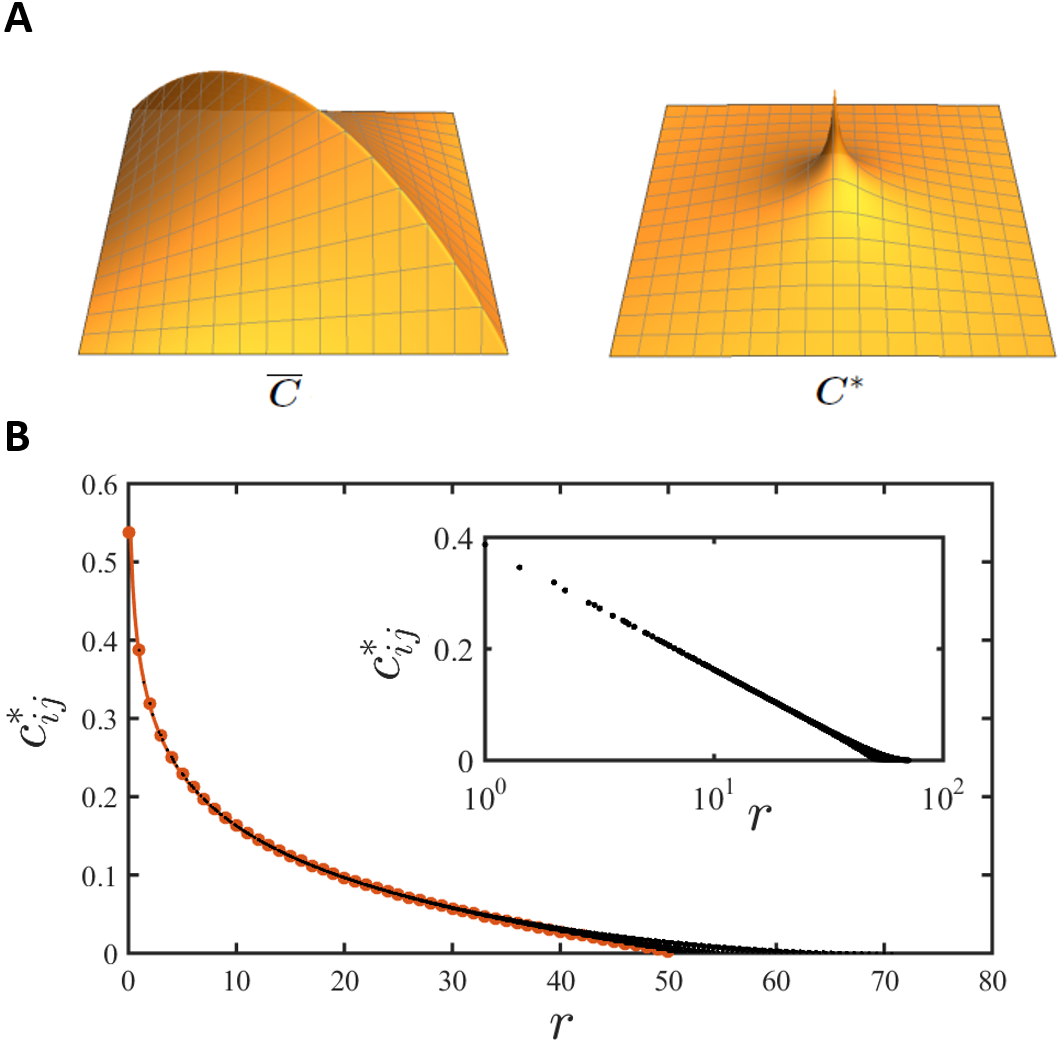}
\caption{A) Profiles of the matrices $\eC$ and $\nC$ in the single activity case. For visual purposes, the discrete data points have been interpolated to a 2D-surface. $\ec_{ij}$ is linear in both indices, resulting in $\omega_{ij}=0$ for $\alpha_i=0$. B) Values of $\nc_{ij}$ versus the distance from the center of the matrix $r=\sqrt{(i-N)^2+(j-N)^2}$. Orange points correspond to the entries at positions (i,N). The inset depicts the same plot in log-linear scale. }
\label{fig:numerC}
\end{figure}

The functional form that we obtain from this approximate analysis accurately describes the actual numerically obtained values of $\nc_{ij}$ far from the boundaries (see Fig.~\ref{fig:numerC}B). Note, the deviations that appear close to the boundaries are due to neglecting the absorbing boundary conditions and not due to the discrete nature of the problem. We can also consider a continuous limit of the problem, in which the chain is replaced by a string, by taking the limit $N\to\infty$, $k\to \infty$, while keeping $k/N=const$. In this limit the discrete diffusion equation is replaced by a continuous one, but the boundary effects still play the same role. We shall return to the continuous limit in Sec.~\ref{subsec4}, where we  discuss more complex networks.

We can use the approximate form of $\nc(r)$ together with  Eq.\eqref{eq:omegalin} to calculate $\omega_{\rm single}(r)$. For large $N$, we can replace $\pdy\nc_{N,N+r}$ with $\partial_r^2 \nc(r)$, to arrive at:
\begin{equation}
\omega_{\rm single}(r)=\frac{\alpha}{T}\frac{1}{\pi r^2}\frac{1}{\sqrt{\det\eCr(r)}}+\mathcal{O}\left(\frac{\alpha^2}{T^2}\right)
\label{eq:omegasin}
\end{equation}
Interestingly, it turns out that $b$, which is in general unknown, does not enter the equation for $\omega(r)$. 

\begin{figure}
\centering
\includegraphics[width=8cm]{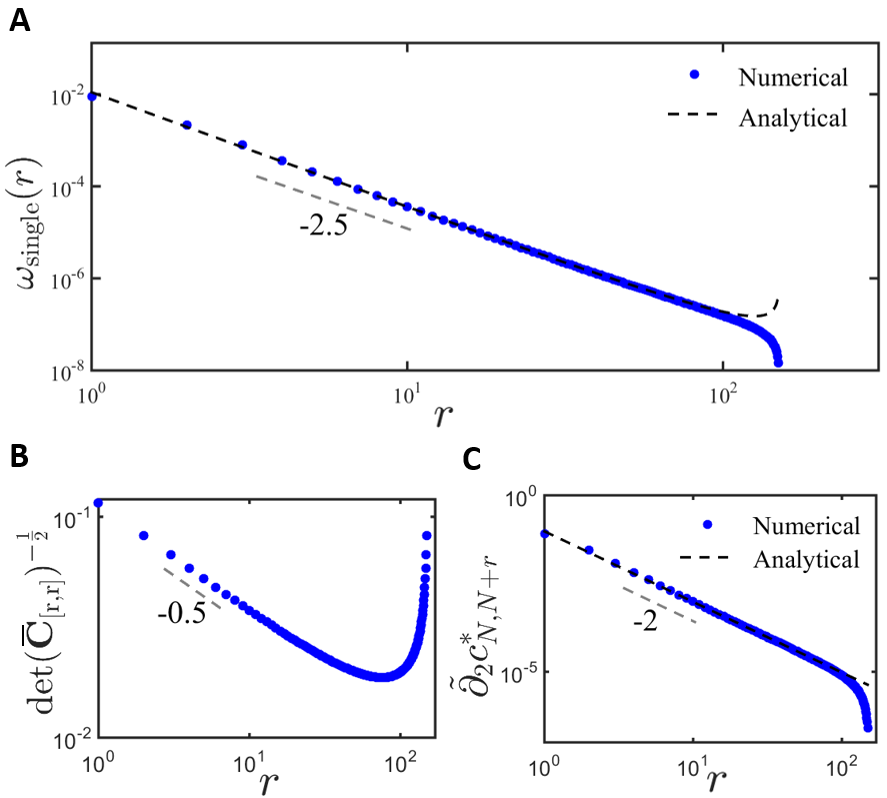}
\caption{ A) Scaling behavior of $\omega_{\rm single}(r)$ as a function of the distance between the beads and comparison with the analytical prediction (dashed line) in Eq.~\eqref{eq:omegasin}. Below, the behaviour of the B) equilibrium and C) non-equilibrium contributions to $\omega_{\rm single}(r)$. The contribution presented in C) coincides with the area enclosing rates. All data points correspond to results obtained by numerically solving Lyapunov equation.
}
\label{fig:1hot}
\end{figure}

To find the equilibrium part $\eC$, we make the following observation. At equilibrium all cycling frequencies $\omega_{ij}$ must vanish, which combined with Eq.~\eqref{eq:omchain} gives: $\pdy\ec_{ij}=0\ \forall_{i\neq j}$. Using Eq.~\eqref{eq:diffec} at point $(i,i)$ and the symmetry of $\eC$ we find $\pdy\ec_{ii}=-1$. In general, we can therefore write the equation for $\eC$ as $\pdy\ec_{ij}=-\delta_{ij}$. Note that this condition is equivalent to a discrete stationary diffusion equation in 1 dimension, with a single source at site $i$ and with absorbing boundary conditions. This implies that $\ec_{ij}$ is linear in both indices and one can easily verify that $\ec_{ij}=\min(i,j)-ij/(2N)$ satisfies Eq.~\eqref{eq:diffec}. Using this solution, we find that $(\det\eCr(r))^{-\frac{1}{2}}=(\frac{1}{2}r(N-r))^{-\frac{1}{2}}\sim r^{-\frac{1}{2}}$ for $r\ll N$. Therefore, $\omega_{\rm single}(r)\sim r^{-\frac{5}{2}}$, as shown together with the numerical results in Fig.~\ref{fig:1hot}.
The good agreement between the numerical and analytical results allows us to conclude that the scaling exponent is determined by the $\ln(r)$-like profile of $\nC$, which in turn is set by the dimensionality of the system. One could in principle find an analytical solution for $\nc(r)$ that accounts for the boundary conditions, but the $\ln(r)$ scaling captures the essential features.

\subsection{Spatially varying activity}
\begin{figure}
\centering
\includegraphics[width=8cm]{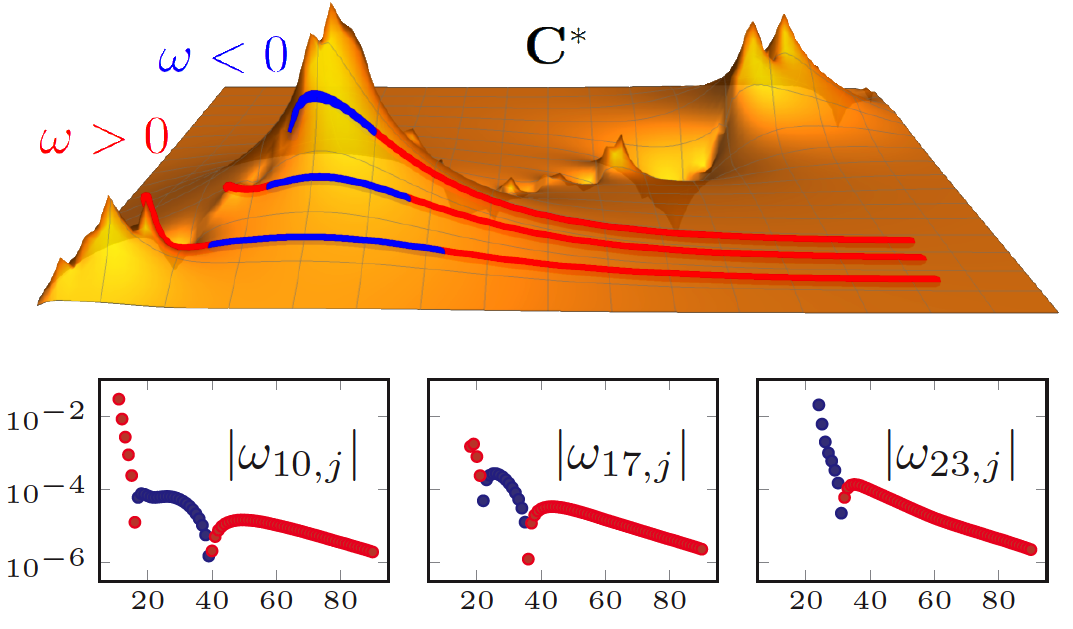}
\caption{Profile of $\nC$ in the case of spatially correlated amplitudes of the active noise, together with the lines $\{(i,j,\nc_{ij})\}_{j=i,\ldots,2N-1}$ for $i=10, 17, 23$. The color of a line indicates the sign of its curvature. The red (blue) points in the plots of $|\omega_{i,j}|$ correspond to positive (negative) cycling frequencies. Here, the profile of $\nC$ is presented for activities $\{\alpha_i-\bar{\alpha}\}$. This transformation of activities is justified in App.~\ref{appendix1}.}
\label{fig:nowypiekny}
\end{figure}

Equipped with the results from the previous section, we now consider a system with spatially varying activity. First, we will further clarify the connection between calculating cycling frequencies and solving a discretized steady-state diffusion equation for the covariance function. To this end, we consider a generic activity distribution $\{ \alpha_i \}$ and plot the corresponding active part of the covariance matrix $\nC$ obtained by solving Eq.~\eqref{eq:diffnc} (see Fig.~\ref{fig:nowypiekny}). From the form of Eq.~\eqref{eq:omchain}, we see that the cycling frequencies $\{\omega_{ij}\}_{j=1,\ldots,2N-1}$ are proportional to the curvature of the line $\{(i,j,\nc_{ij})\}_{j=1,\ldots,2N-1}$, as illustrated by the plots in Fig.~\ref{fig:nowypiekny}.
The connection with the steady-state diffusion equation (see Eq.~\eqref{eq:diffnc}) allows us to understand how a given distribution of activities translates to a particular profile of the $\nC$ matrix and how this in turn determines the behavior of the cycling frequencies. 

In general the amplitudes of the active noise may be spatially correlated. Here, however, we restrict ourselves exclusively to the case of spatially uncorrelated activities, which is valid in the limit of distances larger than the correlation length of the amplitudes. To the ``$i^{th}$" bead we assign a randomly sampled amplitude $\alpha_i$. We assume all $\alpha_i$ to be pairwise independent and identically distributed with distribution $p(\alpha)$. For simplicity we index the beads so that the bead in the center of the system has index $0$. To calculate $\omega(2r)=\omega_{-r,r}$, we need to determine $\pdy \nc_{-r,r}$. For a given activity distribution $\{ \alpha_i\}$ we can exploit the linearity of Eq.~\eqref{eq:diffnc} to obtain the corresponding $\nC(\{ \alpha_i\} )$ as a superposition of single-source solutions. Thus, we can write:
\begin{equation}
\pdy \nc_{-r,r}(\{\alpha_i\} )=\sum_z \pdy \nc_{-r,r}(\{ \alpha_i\delta_{iz} \} )
\label{eq:superposition}
\end{equation}
For beads far enough from the boundary, we approximate $\nC (\{ \alpha_i\delta_{ij}\} )$ by a logarithmic decay centred at $(j,j)$  to obtain:
\begin{align}
\pdy \nc_{-r,r}(\{ \alpha_i\delta_{iz} \} )&=\frac{1}{\pi}\frac{\alpha_z}{\bar{\alpha}} \frac{(r+z)^2-(r-z)^2}{((r+z)^2+(r-z)^2)^2}\nonumber \\
&=\frac{1}{\pi}\frac{\alpha_z}{\bar{\alpha}}\frac{rz}{(r^2+z^2)^2}
\label{eq:partfromz}
\end{align}
Combining Eqs.~\eqref{eq:superposition} and \eqref{eq:partfromz}, we calculate $\langle(\pdy \nc_{-r,r}(\{\alpha_i\} ))^2\rangle$, which is the main factor in the expression for $\langle\omega^2(2r)\rangle$.
\begin{widetext}
\begin{align}
(\pi\bar{\alpha})^2\langle(\pdy \nc_{-r,r}(\{\alpha_i\} ))^2\rangle&=\mean{\left(\sum_{z}\frac{rz\alpha_z}{(r^2+z^2)^2}\right)\left(\sum_{z'}\frac{rz'\alpha_{z'}}{(r^2+z'^2)^2}\right)}=
\sum_{z}\frac{r^2z^2\langle\alpha_z^2\rangle}{(r^2+z^2)^4}+\sum_{z}\frac{rz\mean{\alpha_z}}{(r^2+z^2)^2}\sum_{z'\neq z}\frac{rz'\mean{\alpha_{z'}} }{(r^2+z'^2)^2}\nonumber\\
&=\sum_{z}\frac{r^2z^2(\langle\alpha_z^2\rangle-\mean{\alpha_z}^2)}{(r^2+z^2)^4} \approx \sigma_\alpha^2\int_{-\infty}^{\infty} \frac{r^2z^2dz}{(r^2+z^2)^4}=\frac{\pi\sigma_\alpha^2}{16r^3}
\label{eq:calculate1D}
\end{align}
\end{widetext}
In the second line of this result, we used that $\sum_{z'\neq z}\frac{rz'}{(r^2+z'^2)^2}=-\frac{rz}{(r^2+z^2)^2}$ and approximated the sum by an integral. Evaluating the integral and rescaling $2r\to r$, we arrive at the final result:
\begin{equation}
\langle\omega^2(r)\rangle_\alpha=\frac{\sigma_\alpha^2}{T^2}\frac{1}{2\pi r^3}\frac{1}{\det\eCr(r)}.
\label{eq:scaling}
\end{equation}
Given the asymptotic behavior $1/\det\eCr(r)\sim r^{-1}$ for $r\ll N$ we conclude that in the limit of weak activity $\sqrt{\langle\omega^2(r)\rangle}\sim r^{-2}$ (See Fig.~\ref{fig:omega1D}). Importantly, apart from reproducing the observed exponent of the power-law, our result gives a correct prediction for the prefactor, which contains information about the variance of the active forces.

\section{$d$-dimensional lattices}
\label{sec4}
\begin{figure}
\centering
\includegraphics[width=8cm]{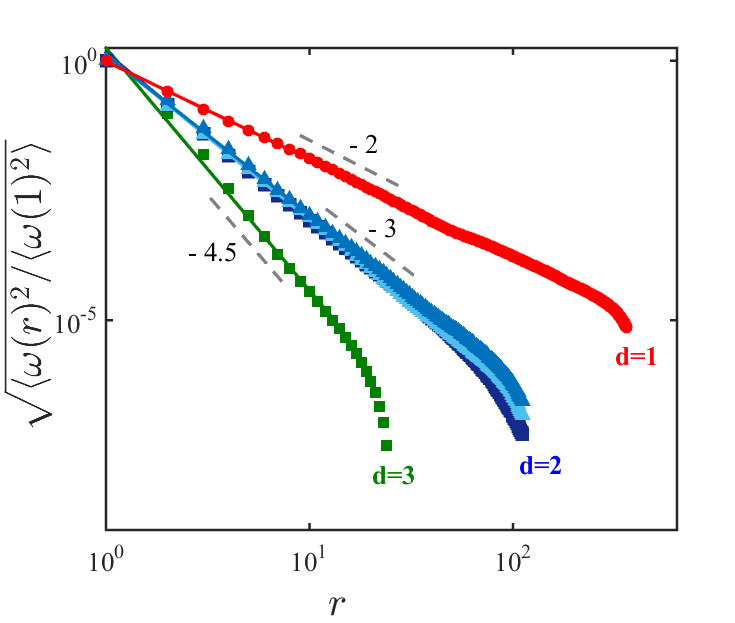}
\caption{Scaling behavior of the cycling frequencies as function of the distance between the beads. The results are obtained for different lattices and a folded Gaussian activity distribution~\cite{Mura2018}. Triangular and square markers represent triangular and square or cubic lattices, respectively. Light (dark) blue represent triangular networks with zero (finite) rest length springs. In all cases  we used $\frac{\bar{\alpha}}{T}=0.15$. For computational convenience we determined the ensemble average by performing a spatial average. All data points correspond to results obtained by numerically solving Lyapunov equation.
}
\label{fig:omega23D}
\end{figure}
\subsection{Cubic lattice}
\label{sec:Cubic lattice}
To explain the origin of the scaling behavior of $\omega(r)$ for multidimensional networks we now focus on the simplest possible case of a $d$-dimensional cubic lattice. Importantly, the calculation presented for this case also provides us with intuition for more complex lattices. 
Let us denote the bead indices corresponding to $d$ independent directions with $n_1,\ldots ,n_d$. We will denote the elements of the covariance matrix $\C$ as $c_{n_1,\ldots,n_d;\nb_1,\ldots,\nb_d}\coloneqq c_{\N,\Nb}$. We assume zero-restlength for the springs, so that the degrees of freedom corresponding to different directions decouple. Therefore, by $\C$ we actually mean the covariance matrix of only these degrees of freedom that correspond to a single chosen direction, for instance the one corresponding to the index $n_1$.

For this particular network, the Lyapunov equation is equivalent to:
\begin{equation}
\label{eq:diff}
\left(\sd\pdni+\sd\pdnib\right)c_{\N,\Nb}=-2d_{\N,\Nb}
\end{equation}
Similarly to the 1-dimensional case, here we recognise a discretized stationary diffusion equation in $2d$ dimensions, with the divergence of the sources given by the elements of $\D$.
For convenience we index the beads such that the one in the middle of the lattice is $(0,\ldots,0)$. 

Our goal is to calculate the cycling frequency $\omega_{r,\ldots,r;-r,\ldots,-r}=\omega(2\sqrt{d}r)$. Here we consider a particular case of relative position of the beads with respect to the principal directions of the lattice. It turns out, however, that the final result depends only on the distance between the beads. From Eq.~\eqref{eq:torque}, we find that:
\begin{equation}
\label{eq:omegalind}
\omega_{r,\ldots,r;-r,\ldots,-r}\approx \frac{(\bar{\alpha}/T)}{\sqrt{\det\eCr}}\sd\left. \pdni \nc_{\N,\Nb}\right|_{r,\ldots,r;-r,\ldots,-r}
\end{equation}
Following the procedure used for the 1-dimensional chain, we begin with finding the solution to a single source problem with one active bead at site $(0,\ldots,0)$. As before, we will then use this solution as a Green's function for our diffusion problem with a generic activity distribution.

Taking a continuous limit of the diffusion equation and neglecting the boundary conditions, we expect that $\partial_r c(r)\sim 1/r^{2d-1}$ and consequently $c(r)\sim 1/r^{2d-2}$, where $r$ is the distance from the center of the $2d$-dimensional covariance matrix. Therefore, for a single active bead at site $(0,\ldots,0)$, we obtain:
\begin{equation}
\nc_{\N,\Nb}=a_d\left( \sd n_i^2 +\sd \nb_i^2 \right)^{-(d-1)}\mkern-18mu =a_d\left( \N^2 +\Nb^2 \right)^{-(d-1)}
\end{equation}
The constant $a_d=(d-2)!/(2\pi^d)$ can be obtained from the divergence theorem, as we did in the 1-dimensional case. 

The contribution to $\omega_{\N,\Nb}$ from a single activity at site $(0,\ldots,0)$ is then given by (App.~\ref{appendix1}): 
\begin{equation}
\label{eq:partials}
\sd\partial_{n_i}^2\nc_{\N,\Nb}=2 d(d-1)a_d\frac{\N^2-\Nb^2}{(\N^2+\Nb^2)^{d+1}}
\end{equation}
Performing calculations analogous to those for the 1~-~dimensional chain, we arrive at (App.~\ref{appendix1})
\begin{align}
\mean{\omega_{d=2}^2(r)}_\alpha&=\frac{\sigma_\alpha^2}{T^2}\frac{8}{5\pi^3r^6}\frac{1}{\det\eCr}\\
\mean{\omega_{d=3}^2(r)}_\alpha&=\frac{\sigma_\alpha^2}{T^2}\frac{27}{8\pi^4r^9}\frac{1}{\det\eCr}
\end{align}
Importantly, we obtain exactly the same results when considering different directions across the lattice, such as $\omega_{(r,0,\ldots,0),(-r,0,\ldots,0)}$. In general, for a $d$-dimensional lattice we expect:
\begin{align}
\partial_r^2c(r)&\sim r^{-2d}\\
\omega^2_{{\rm single},d}(r)&\sim r^{-4d}/\det\eCr(r)\\
\langle\omega^2_d(r)\rangle_\alpha&\sim r^{-3d}/\det\eCr(r)
\end{align}

For completeness we investigate the behavior of $\det\eCr(r)$ for different dimensions. At equilibrium all cycling frequencies vanish, leading to:
\begin{equation}
\sd\pdni \ec_{\N,\Nb}=0\quad \forall_{\N\neq \Nb}
\label{eq:cond1}
\end{equation}
For all points on the diagonal of the covariance matrix $(\N,\N)=(n_1,\ldots,n_d;n_1,\ldots,n_d)$ the diffusion equation (see Eq.~\eqref{eq:diff}) reads
\begin{equation}
\sd\pdni\ec_{\N,\N}+\sd\pdnib\ec_{\N,\N}=-2
\label{eq:cond15}
\end{equation}
Using the symmetry of the system, we conclude that the two sums in Eq.~\eqref{eq:cond15} are equal, which together with Eq.~\eqref{eq:cond1} imply that
\begin{equation}
(\sd \pdni)\ec_{\N,\Nb}=-\delta_{\N,\Nb}
\label{eq:cond2}
\end{equation}
for all points $(\N,\Nb)$. This result can be interpreted in the following way: for a given $(\nb_1,\ldots,\nb_d)$, $\ec_{\N,\Nb}$ as a function of $(n_1,\ldots,n_d)$ is a solution to a $d$-dimensional discretised stationary diffusion equation with a single source at position $(\nb_1,\ldots,\nb_d)$, and with absorbing boundary conditions. Note, there is an interesting symmetry of the diffusion equation implied by the symmetry $\ec_{\N,\Nb}=\ec_{\Nb,\N}$: the solution at point $\N$ from a source at point $\Nb$ is equal to the solution at point $\Nb$ from a source at point $\N$. While this property of the diffusion equation would be obvious in an infinite space, it surprisingly holds also in the presence of absorbing boundaries. 

It can further be shown that
\begin{align}
\ec_{\N,\Nb}\sim\ln[(n_1-\nb_1)^2+(n_2-\nb_2)^2]\quad \mathrm{for}\ d=2\nonumber\\
\ec_{\N,\Nb}\sim \left( \sd (n_i-\nb_i)^2 \right)^{-\left( \frac{d-2}{2} \right)}\ \mathrm{for}\ d> 2
\label{eq:eCscaling}
\end{align}
This result can also be understood using a simple dimensionality argument: a diffusion problem in  $d$ dimensions with a source forming a $d_s$-dimensional plane can be mapped to a $(d-d_s)$-dimensional diffusion problem with a point source. In our case we are dealing with a diffusion problem in a $2d$-dimensional space, with a $d$-dimensional source. Reducing the $2d$-dimensional problem to a point source problem in $d$ dimensions, we arrive exactly at Eq.~\eqref{eq:eCscaling}.
From this equation we conclude that for dimensions $d\geq2$, the diagonal terms of $\eCr(r)$ strongly dominate over the off-diagonal ones. In fact one can verify that for dimensions $d\geq 2$ and for systems large enough $\det\eCr(r)$ depends on $r$ only weakly and does not influence the scaling behavior of $\omega^2(r)$ anymore (See Fig.~\ref{fig:omega23D}). This is a consequence of the shorter range of elastic interactions in higher dimensions.

It is important to note here that, as discussed after introducing Eq.~\eqref{eq:omegalin}, the area enclosing rates do not depend on $\det\Cr$. This allows us to perform calculations analogous to the ones presented in this section, without assuming the limit of weak activities. As a result we predict a scaling $\sigma_\alpha r^{-3d/2}$ for the area enclosing rates even for high amplitudes of the active noise.

\subsection{Generic lattices}
\label{subsec4}
\begin{figure}
\centering
\includegraphics[width=8cm]{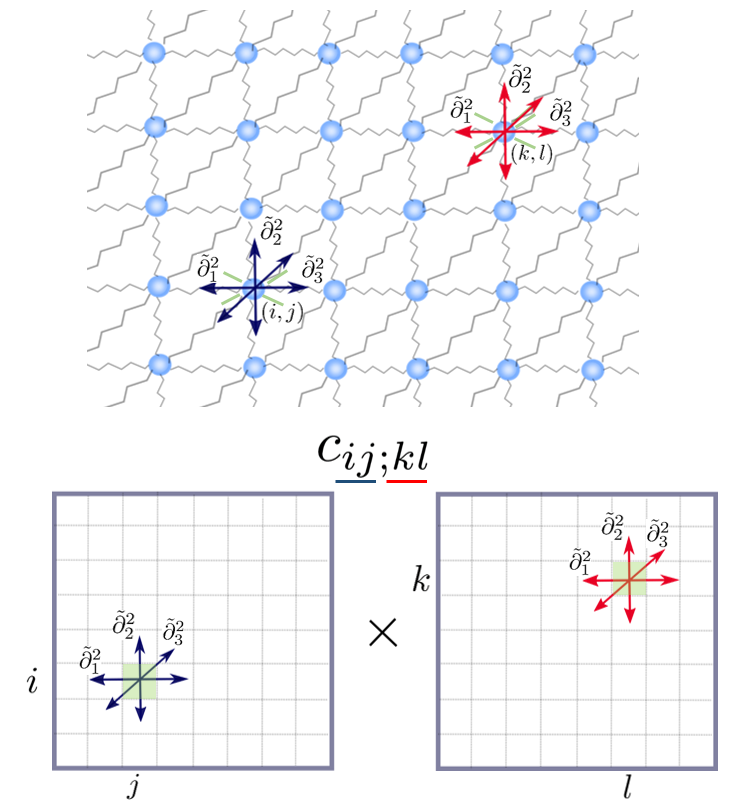}
\caption{An exemplary lattice for which the Lyaunov equation can be interpreted as a diffusion equation with non-isotropic diffusion. The presented lattice is equivalent to the triangular lattice, if we consider them in the zero-rest length case.}
\label{fig:lattice2}
\end{figure}

In Sec.~\ref{sec:Cubic lattice} we investigated the simplest possible case of a $d$-dimensional zero-rest length cubic lattice $\mathcal{L}^d$. For such systems the Lyapunov equation for the covariance matrix could be viewed as a discretized steady state diffusion equation defined on a space $\mathcal{L}^d~\times ~\mathcal{L}^d~\sim ~\mathcal{L}^{2d}$. For instance, for a 2-dimensional square lattice we had to solve a diffusion equation in a 4-dimensional cube. A natural question is how general the connection is between the Lyapunov equation and diffusion equations. It turns out that for many zero-rest length lattices there is simple procedure for translating a particular lattice structure to a corresponding diffusion equation for the covariance. The condition which allows us to identify the terms appearing in the Lyapunov equation with second derivatives, as in Eq.~\eqref{eq:diffnc}, is that at all sites of the lattice a spring pointing in one direction is accompanied by a spring pointing in the opposite direction. If this is the case we can directly read out the diffusion equation from the structure of the lattice, as illustrated in Fig.~\ref{fig:lattice2}. Each such pair of springs gives rise to diffusive terms in the corresponding directions, with diffusion constant proportional to the spring constant.

In the case of finite rest length elastic networks with linearized forces, the same condition  allows us to write the Lyapunov equation as a discretized second order partial differential equation for the covariance. Importantly, the displacements in $x$- and $y$-directions are no longer decoupled and one has to solve a differential equation for three different covariances: $c^{xx}$, $c^{xy}$, $c^{yy}$. An example of such an equation for a 2-dimensional triangular lattice is included in App.~\ref{appendix2}. Importantly, the structure of the network determines not only the equation for the covariance matrix but also the cycling frequencies according to Eq.~\eqref{eq:torque}.

For a wide range of networks, including randomly diluted networks \cite{Gnesotto2018}, the condition given above is not satisfied and there is no straightforward way of translating the Lyapunov equation to a continuous diffusion-like equation for the covariance. Nevertheless, for a given network $\mathcal{G}$, which can be thought of as a graph, we can still interpret the Lyapunov equation as a Poisson equation on a graph $\mathcal{G}\times\mathcal{G}$ and relate the cycling frequency between a pair of degrees of freedom to the covariance flux through a corresponding vertex of $\mathcal{G}\times\mathcal{G}$. The theory of graph Laplacians, introduced by Kirchhoff in his study of the properties of resistor networks, has found applications in elasticity theory, graph theory, and computer science~\cite{Mohar1991,Schaeffer2007,Fisher1966,Kirchhoff1847}.

Our numerical calculations reveal that the exponents of the power-laws for the cycling frequencies observed for various lattices are set by the dimensionality of the system and are independent of the detailed structure of the lattice~\cite{Mura2018}. Heuristically, this can be understood as follows: any $d$-dimensional rigid network, with a given average coordination number can be seen as an approximation to a continuous $d$-dimensional elastic medium. For such an elastic medium, a continuous diffusion equation, as the one we used to study $d$-dimensional cubic lattices, would be an exact equation for the covariance field, and we conjecture that the cycling frequencies for a continuous medium can be obtained by taking appropriate limits of our results for a discrete system. Note that a study of the cycling frequencies directly at the level of a continuous system would require introducing the Fokker-Planck equation for fields and make the analysis considerably more difficult. Since our results presented for the $d$-dimensional cubic lattice should coincide with the results for a $d$-dimensional continuous medium, we argue that our analytical calculation captures the essential origins of the power-law phenomenon for all lattices that approximate a continuous medium well.

\section{Conclusions}
Here we considered a simple model of an internally driven elastic assembly. Using this model, we investigated the properties of cycling frequencies - a two-point measure of non-equilibrium dynamics, which can be used in experimental and theoretical studies of active systems. We discussed how to relate the cycling frequencies to other commonly used non-equilibrium measures, such as the area enclosing rates or the reduced entropy production rate. Furthermore, based on our analytical approach, reinforced with numerical results, we predict that  the cycling frequencies follow a power-law as a function of distance between the two probes in an elastic network. The exponent of the power-law depends on the dimensionality of the system, but not on the detailed structure of the network. In the case of a random spatial distribution of activities, we showed that the mean cycling frequencies $\sqrt{\langle\omega^2(r)\rangle}$ are proportional to the standard deviation of the intensities of the active noise $\sigma_\alpha$. Interestingly, the case of a single-site activity gives a different exponent than the one with randomly distributed activities.

For more complex activity distributions, the connection between the Lyapunov equation and a diffusion equation, that we presented in sections~\ref{sec3} and~\ref{sec4}, provides some intuition for how the cycling frequencies in a system depend on the structure of the active noise. Since the diffusive terms in the Lyapunov equation originate solely from the structure of the lattice, we expect that a similar framework can be used to study the behavior of the cycling frequencies for more complex distributions of the active noise, which include spatial correlations~\cite{Mura2019}.

The analytical approach we developed aims at finding a mapping between the properties of the active noise and two-point non-equilibrium measures. Based on the results we obtained and their robustness to the detailed structure of a network, we argue that the cycling frequencies and the area enclosing rates are promising tools for studying the nature of the stochastic driving in an active elastic assembly. Our approach can be tested on reconstituted acto-myosin networks~\cite{Liu2006,Koenderink2009,Alvarado2017} and other noise-driven biological~\cite{Weber,Betz2009,Turlier2016} or synthetic systems~\cite{Boekhoven2015,Palacci2013,Bertrand2012}, which can be well approximated by an elastic assembly at steady-state. Such  experiments could be performed in chromosomes, membranes or tissues, using embedded colloidal particles, or fluorescently tagged cellular organelles.

\begin{acknowledgments}
We thank F.~Gnesotto, B.~Remlein, and P.~Ronceray for many stimulating discussions.
This work was supported by the German Excellence Initiative via
 the program NanoSystems Initiative Munich (NIM), the Graduate School of Quantitative Biosciences Munich (QBM),  and was funded by the Deutsche Forschungsgemeinshaft (DFG, German Research Foundation) - 418389167.
 \end{acknowledgments}

\bibliography{Newbiblio}

\begin{thebibliography}{55}%
\makeatletter
\providecommand \@ifxundefined [1]{%
 \@ifx{#1\undefined}
}%
\providecommand \@ifnum [1]{%
 \ifnum #1\expandafter \@firstoftwo
 \else \expandafter \@secondoftwo
 \fi
}%
\providecommand \@ifx [1]{%
 \ifx #1\expandafter \@firstoftwo
 \else \expandafter \@secondoftwo
 \fi
}%
\providecommand \natexlab [1]{#1}%
\providecommand \enquote  [1]{``#1''}%
\providecommand \bibnamefont  [1]{#1}%
\providecommand \bibfnamefont [1]{#1}%
\providecommand \citenamefont [1]{#1}%
\providecommand \href@noop [0]{\@secondoftwo}%
\providecommand \href [0]{\begingroup \@sanitize@url \@href}%
\providecommand \@href[1]{\@@startlink{#1}\@@href}%
\providecommand \@@href[1]{\endgroup#1\@@endlink}%
\providecommand \@sanitize@url [0]{\catcode `\\12\catcode `\$12\catcode
  `\&12\catcode `\#12\catcode `\^12\catcode `\_12\catcode `\%12\relax}%
\providecommand \@@startlink[1]{}%
\providecommand \@@endlink[0]{}%
\providecommand \url  [0]{\begingroup\@sanitize@url \@url }%
\providecommand \@url [1]{\endgroup\@href {#1}{\urlprefix }}%
\providecommand \urlprefix  [0]{URL }%
\providecommand \Eprint [0]{\href }%
\providecommand \doibase [0]{http://dx.doi.org/}%
\providecommand \selectlanguage [0]{\@gobble}%
\providecommand \bibinfo  [0]{\@secondoftwo}%
\providecommand \bibfield  [0]{\@secondoftwo}%
\providecommand \translation [1]{[#1]}%
\providecommand \BibitemOpen [0]{}%
\providecommand \bibitemStop [0]{}%
\providecommand \bibitemNoStop [0]{.\EOS\space}%
\providecommand \EOS [0]{\spacefactor3000\relax}%
\providecommand \BibitemShut  [1]{\csname bibitem#1\endcsname}%
\let\auto@bib@innerbib\@empty
\bibitem [{\citenamefont {Fodor}\ and\ \citenamefont {{Cristina
  Marchetti}}(2018)}]{Fodorreview}%
  \BibitemOpen
  \bibfield  {author} {\bibinfo {author} {\bibfnamefont {{\'{E}}.}~\bibnamefont
  {Fodor}}\ and\ \bibinfo {author} {\bibfnamefont {M.}~\bibnamefont {{Cristina
  Marchetti}}},\ }\href {\doibase 10.1016/j.physa.2017.12.137} {\bibfield
  {journal} {\bibinfo  {journal} {Phys. A Stat. Mech. its Appl.}\ }\textbf
  {\bibinfo {volume} {504}},\ \bibinfo {pages} {106} (\bibinfo {year}
  {2018})},\ \Eprint {http://arxiv.org/abs/1708.08652} {arXiv:1708.08652}
  \BibitemShut {NoStop}%
\bibitem [{\citenamefont {J{\"{u}}licher}\ \emph {et~al.}(2007)\citenamefont
  {J{\"{u}}licher}, \citenamefont {Kruse}, \citenamefont {Prost},\ and\
  \citenamefont {Joanny}}]{Jülicherreview}%
  \BibitemOpen
  \bibfield  {author} {\bibinfo {author} {\bibfnamefont {F.}~\bibnamefont
  {J{\"{u}}licher}}, \bibinfo {author} {\bibfnamefont {K.}~\bibnamefont
  {Kruse}}, \bibinfo {author} {\bibfnamefont {J.}~\bibnamefont {Prost}}, \ and\
  \bibinfo {author} {\bibfnamefont {J.~F.}\ \bibnamefont {Joanny}},\ }\href
  {\doibase 10.1016/j.physrep.2007.02.018} {\bibfield  {journal} {\bibinfo
  {journal} {Phys. Rep.}\ }\textbf {\bibinfo {volume} {449}},\ \bibinfo {pages}
  {3} (\bibinfo {year} {2007})}\BibitemShut {NoStop}%
\bibitem [{\citenamefont {Marchetti}\ \emph {et~al.}(2013)\citenamefont
  {Marchetti}, \citenamefont {Joanny}, \citenamefont {Ramaswamy}, \citenamefont
  {Liverpool}, \citenamefont {Prost}, \citenamefont {Rao},\ and\ \citenamefont
  {Simha}}]{Marchettireview}%
  \BibitemOpen
  \bibfield  {author} {\bibinfo {author} {\bibfnamefont {M.~C.}\ \bibnamefont
  {Marchetti}}, \bibinfo {author} {\bibfnamefont {J.~F.}\ \bibnamefont
  {Joanny}}, \bibinfo {author} {\bibfnamefont {S.}~\bibnamefont {Ramaswamy}},
  \bibinfo {author} {\bibfnamefont {T.~B.}\ \bibnamefont {Liverpool}}, \bibinfo
  {author} {\bibfnamefont {J.}~\bibnamefont {Prost}}, \bibinfo {author}
  {\bibfnamefont {M.}~\bibnamefont {Rao}}, \ and\ \bibinfo {author}
  {\bibfnamefont {R.~A.}\ \bibnamefont {Simha}},\ }\href {\doibase
  10.1103/RevModPhys.85.1143} {\bibfield  {journal} {\bibinfo  {journal} {Rev.
  Mod. Phys.}\ }\textbf {\bibinfo {volume} {85}},\ \bibinfo {pages} {1143}
  (\bibinfo {year} {2013})},\ \Eprint {http://arxiv.org/abs/1207.2929}
  {arXiv:1207.2929} \BibitemShut {NoStop}%
\bibitem [{\citenamefont {Filella}\ \emph {et~al.}(2018)\citenamefont
  {Filella}, \citenamefont {Nadal}, \citenamefont {Sire}, \citenamefont
  {Kanso},\ and\ \citenamefont {Eloy}}]{Filella2018}%
  \BibitemOpen
  \bibfield  {author} {\bibinfo {author} {\bibfnamefont {A.}~\bibnamefont
  {Filella}}, \bibinfo {author} {\bibfnamefont {F.}~\bibnamefont {Nadal}},
  \bibinfo {author} {\bibfnamefont {C.}~\bibnamefont {Sire}}, \bibinfo {author}
  {\bibfnamefont {E.}~\bibnamefont {Kanso}}, \ and\ \bibinfo {author}
  {\bibfnamefont {C.}~\bibnamefont {Eloy}},\ }\href {\doibase
  10.1103/PhysRevLett.120.198101} {\bibfield  {journal} {\bibinfo  {journal}
  {Phys. Rev. Lett.}\ }\textbf {\bibinfo {volume} {120}},\ \bibinfo {pages}
  {35} (\bibinfo {year} {2018})},\ \Eprint {http://arxiv.org/abs/1705.07821}
  {arXiv:1705.07821} \BibitemShut {NoStop}%
\bibitem [{\citenamefont {Ballerini}\ \emph {et~al.}(2008)\citenamefont
  {Ballerini}, \citenamefont {Cabibbo}, \citenamefont {Candelier},
  \citenamefont {Cavagna}, \citenamefont {Cisbani}, \citenamefont {Giardina},
  \citenamefont {Lecomte}, \citenamefont {Orlandi}, \citenamefont {Parisi},
  \citenamefont {Procaccini}, \citenamefont {Viale},\ and\ \citenamefont
  {Zdravkovic}}]{Ballerini2008}%
  \BibitemOpen
  \bibfield  {author} {\bibinfo {author} {\bibfnamefont {M.}~\bibnamefont
  {Ballerini}}, \bibinfo {author} {\bibfnamefont {N.}~\bibnamefont {Cabibbo}},
  \bibinfo {author} {\bibfnamefont {R.}~\bibnamefont {Candelier}}, \bibinfo
  {author} {\bibfnamefont {A.}~\bibnamefont {Cavagna}}, \bibinfo {author}
  {\bibfnamefont {E.}~\bibnamefont {Cisbani}}, \bibinfo {author} {\bibfnamefont
  {I.}~\bibnamefont {Giardina}}, \bibinfo {author} {\bibfnamefont
  {V.}~\bibnamefont {Lecomte}}, \bibinfo {author} {\bibfnamefont
  {A.}~\bibnamefont {Orlandi}}, \bibinfo {author} {\bibfnamefont
  {G.}~\bibnamefont {Parisi}}, \bibinfo {author} {\bibfnamefont
  {A.}~\bibnamefont {Procaccini}}, \bibinfo {author} {\bibfnamefont
  {M.}~\bibnamefont {Viale}}, \ and\ \bibinfo {author} {\bibfnamefont
  {V.}~\bibnamefont {Zdravkovic}},\ }\href {\doibase 10.1073/pnas.0711437105}
  {\bibfield  {journal} {\bibinfo  {journal} {Proc. Natl. Acad. Sci. U.S.A.}\
  }\textbf {\bibinfo {volume} {105}},\ \bibinfo {pages} {1232} (\bibinfo {year}
  {2008})},\ \Eprint {http://arxiv.org/abs/0709.1916} {arXiv:0709.1916}
  \BibitemShut {NoStop}%
\bibitem [{\citenamefont {Damton}\ \emph {et~al.}(2010)\citenamefont {Damton},
  \citenamefont {Turner}, \citenamefont {Rojevsky},\ and\ \citenamefont
  {Berg}}]{Damton2010}%
  \BibitemOpen
  \bibfield  {author} {\bibinfo {author} {\bibfnamefont {N.~C.}\ \bibnamefont
  {Damton}}, \bibinfo {author} {\bibfnamefont {L.}~\bibnamefont {Turner}},
  \bibinfo {author} {\bibfnamefont {S.}~\bibnamefont {Rojevsky}}, \ and\
  \bibinfo {author} {\bibfnamefont {H.~C.}\ \bibnamefont {Berg}},\ }\href
  {\doibase 10.1016/j.bpj.2010.01.053} {\bibfield  {journal} {\bibinfo
  {journal} {Biophys. J.}\ }\textbf {\bibinfo {volume} {98}},\ \bibinfo {pages}
  {2082} (\bibinfo {year} {2010})}\BibitemShut {NoStop}%
\bibitem [{\citenamefont {Thutupalli}\ \emph {et~al.}(2015)\citenamefont
  {Thutupalli}, \citenamefont {Sun}, \citenamefont {Bunyak}, \citenamefont
  {Palaniappan},\ and\ \citenamefont {Shaevitz}}]{Thutupalli2015a}%
  \BibitemOpen
  \bibfield  {author} {\bibinfo {author} {\bibfnamefont {S.}~\bibnamefont
  {Thutupalli}}, \bibinfo {author} {\bibfnamefont {M.}~\bibnamefont {Sun}},
  \bibinfo {author} {\bibfnamefont {F.}~\bibnamefont {Bunyak}}, \bibinfo
  {author} {\bibfnamefont {K.}~\bibnamefont {Palaniappan}}, \ and\ \bibinfo
  {author} {\bibfnamefont {J.~W.}\ \bibnamefont {Shaevitz}},\ }\href {\doibase
  10.1098/rsif.2015.0049} {\bibfield  {journal} {\bibinfo  {journal} {J. R.
  Soc. Interface}\ }\textbf {\bibinfo {volume} {12}} (\bibinfo {year} {2015}),\
  10.1098/rsif.2015.0049},\ \Eprint {http://arxiv.org/abs/1410.7230}
  {arXiv:1410.7230} \BibitemShut {NoStop}%
\bibitem [{\citenamefont {MacKintosh}\ and\ \citenamefont
  {Schmidt}(2010)}]{mackintosh2010}%
  \BibitemOpen
  \bibfield  {author} {\bibinfo {author} {\bibfnamefont {F.~C.}\ \bibnamefont
  {MacKintosh}}\ and\ \bibinfo {author} {\bibfnamefont {C.~F.}\ \bibnamefont
  {Schmidt}},\ }\href {\doibase 10.1016/j.ceb.2010.01.002} {\bibfield
  {journal} {\bibinfo  {journal} {Curr. Opin. Cell Biol.}\ }\textbf {\bibinfo
  {volume} {22}},\ \bibinfo {pages} {29} (\bibinfo {year} {2010})}\BibitemShut
  {NoStop}%
\bibitem [{\citenamefont {Gnesotto}\ \emph
  {et~al.}(2018{\natexlab{a}})\citenamefont {Gnesotto}, \citenamefont {Mura},
  \citenamefont {Gladrow},\ and\ \citenamefont {Broedersz}}]{Gnesottoreview}%
  \BibitemOpen
  \bibfield  {author} {\bibinfo {author} {\bibfnamefont {F.~S.}\ \bibnamefont
  {Gnesotto}}, \bibinfo {author} {\bibfnamefont {F.}~\bibnamefont {Mura}},
  \bibinfo {author} {\bibfnamefont {J.}~\bibnamefont {Gladrow}}, \ and\
  \bibinfo {author} {\bibfnamefont {C.~P.}\ \bibnamefont {Broedersz}},\ }\href
  {http://stacks.iop.org/0034-4885/81/i=6/a=066601?key=crossref.4831a190a4802ca2c9e13c2201982b36}
  {\bibfield  {journal} {\bibinfo  {journal} {Reports Prog. Phys.}\ }\textbf
  {\bibinfo {volume} {81}},\ \bibinfo {pages} {066601} (\bibinfo {year}
  {2018}{\natexlab{a}})},\ \Eprint {http://arxiv.org/abs/1710.03456}
  {arXiv:1710.03456} \BibitemShut {NoStop}%
\bibitem [{\citenamefont {Weber}\ \emph {et~al.}(2012)\citenamefont {Weber},
  \citenamefont {Spakowitz},\ and\ \citenamefont {Theriot}}]{Weber}%
  \BibitemOpen
  \bibfield  {author} {\bibinfo {author} {\bibfnamefont {S.~C.}\ \bibnamefont
  {Weber}}, \bibinfo {author} {\bibfnamefont {A.~J.}\ \bibnamefont
  {Spakowitz}}, \ and\ \bibinfo {author} {\bibfnamefont {J.~A.}\ \bibnamefont
  {Theriot}},\ }\href {\doibase 10.1073/pnas.1119505109} {\bibfield  {journal}
  {\bibinfo  {journal} {Proc. Natl. Acad. Sci. U.S.A.}\ }\textbf {\bibinfo
  {volume} {109}},\ \bibinfo {pages} {7338} (\bibinfo {year}
  {2012})}\BibitemShut {NoStop}%
\bibitem [{\citenamefont {Lau}\ \emph {et~al.}(2003)\citenamefont {Lau},
  \citenamefont {Hoffman}, \citenamefont {Davies}, \citenamefont {Crocker},\
  and\ \citenamefont {Lubensky}}]{Lau2003}%
  \BibitemOpen
  \bibfield  {author} {\bibinfo {author} {\bibfnamefont {A.~W.~C.}\
  \bibnamefont {Lau}}, \bibinfo {author} {\bibfnamefont {B.~D.}\ \bibnamefont
  {Hoffman}}, \bibinfo {author} {\bibfnamefont {A.}~\bibnamefont {Davies}},
  \bibinfo {author} {\bibfnamefont {J.~C.}\ \bibnamefont {Crocker}}, \ and\
  \bibinfo {author} {\bibfnamefont {T.~C.}\ \bibnamefont {Lubensky}},\ }\href
  {\doibase 10.1103/PhysRevLett.91.198101} {\bibfield  {journal} {\bibinfo
  {journal} {Phys. Rev. Lett.}\ }\textbf {\bibinfo {volume} {91}},\ \bibinfo
  {pages} {198101} (\bibinfo {year} {2003})}\BibitemShut {NoStop}%
\bibitem [{\citenamefont {Guo}\ \emph {et~al.}(2014)\citenamefont {Guo},
  \citenamefont {Ehrlicher}, \citenamefont {Jensen}, \citenamefont {Renz},
  \citenamefont {Moore}, \citenamefont {Goldman}, \citenamefont
  {Lippincott-Schwartz}, \citenamefont {Mackintosh},\ and\ \citenamefont
  {Weitz}}]{Guo2014}%
  \BibitemOpen
  \bibfield  {author} {\bibinfo {author} {\bibfnamefont {M.}~\bibnamefont
  {Guo}}, \bibinfo {author} {\bibfnamefont {A.~J.}\ \bibnamefont {Ehrlicher}},
  \bibinfo {author} {\bibfnamefont {M.~H.}\ \bibnamefont {Jensen}}, \bibinfo
  {author} {\bibfnamefont {M.}~\bibnamefont {Renz}}, \bibinfo {author}
  {\bibfnamefont {J.~R.}\ \bibnamefont {Moore}}, \bibinfo {author}
  {\bibfnamefont {R.~D.}\ \bibnamefont {Goldman}}, \bibinfo {author}
  {\bibfnamefont {J.}~\bibnamefont {Lippincott-Schwartz}}, \bibinfo {author}
  {\bibfnamefont {F.~C.}\ \bibnamefont {Mackintosh}}, \ and\ \bibinfo {author}
  {\bibfnamefont {D.~A.}\ \bibnamefont {Weitz}},\ }\href@noop {} {\bibfield
  {journal} {\bibinfo  {journal} {Cell}\ }\textbf {\bibinfo {volume} {158}},\
  \bibinfo {pages} {822} (\bibinfo {year} {2014})}\BibitemShut {NoStop}%
\bibitem [{\citenamefont {Fakhri}\ \emph {et~al.}(2014)\citenamefont {Fakhri},
  \citenamefont {Wessel}, \citenamefont {Willms}, \citenamefont {Pasquali},
  \citenamefont {Klopfenstein}, \citenamefont {MacKintosh},\ and\ \citenamefont
  {Schmidt}}]{Fakhri2014}%
  \BibitemOpen
  \bibfield  {author} {\bibinfo {author} {\bibfnamefont {N.}~\bibnamefont
  {Fakhri}}, \bibinfo {author} {\bibfnamefont {A.~D.}\ \bibnamefont {Wessel}},
  \bibinfo {author} {\bibfnamefont {C.}~\bibnamefont {Willms}}, \bibinfo
  {author} {\bibfnamefont {M.}~\bibnamefont {Pasquali}}, \bibinfo {author}
  {\bibfnamefont {D.~R.}\ \bibnamefont {Klopfenstein}}, \bibinfo {author}
  {\bibfnamefont {F.~C.}\ \bibnamefont {MacKintosh}}, \ and\ \bibinfo {author}
  {\bibfnamefont {C.~F.}\ \bibnamefont {Schmidt}},\ }\href@noop {} {\bibfield
  {journal} {\bibinfo  {journal} {Science.}\ }\textbf {\bibinfo {volume}
  {344}},\ \bibinfo {pages} {1031} (\bibinfo {year} {2014})}\BibitemShut
  {NoStop}%
\bibitem [{\citenamefont {Brangwynne}\ \emph {et~al.}(2008)\citenamefont
  {Brangwynne}, \citenamefont {Koenderink}, \citenamefont {MacKintosh},\ and\
  \citenamefont {Weitz}}]{Brangwynne2008}%
  \BibitemOpen
  \bibfield  {author} {\bibinfo {author} {\bibfnamefont {C.~P.}\ \bibnamefont
  {Brangwynne}}, \bibinfo {author} {\bibfnamefont {G.~H.}\ \bibnamefont
  {Koenderink}}, \bibinfo {author} {\bibfnamefont {F.~C.}\ \bibnamefont
  {MacKintosh}}, \ and\ \bibinfo {author} {\bibfnamefont {D.~A.}\ \bibnamefont
  {Weitz}},\ }\href {\doibase 10.1083/jcb.200806149} {\bibfield  {journal}
  {\bibinfo  {journal} {J. Cell Biol.}\ }\textbf {\bibinfo {volume} {183}},\
  \bibinfo {pages} {583} (\bibinfo {year} {2008})}\BibitemShut {NoStop}%
\bibitem [{\citenamefont {Betz}\ \emph {et~al.}(2009)\citenamefont {Betz},
  \citenamefont {Lenz}, \citenamefont {Joanny},\ and\ \citenamefont
  {Sykes}}]{Betz2009}%
  \BibitemOpen
  \bibfield  {author} {\bibinfo {author} {\bibfnamefont {T.}~\bibnamefont
  {Betz}}, \bibinfo {author} {\bibfnamefont {M.}~\bibnamefont {Lenz}}, \bibinfo
  {author} {\bibfnamefont {J.-F.}\ \bibnamefont {Joanny}}, \ and\ \bibinfo
  {author} {\bibfnamefont {C.}~\bibnamefont {Sykes}},\ }\href {\doibase
  10.1073/pnas.0904614106} {\bibfield  {journal} {\bibinfo  {journal} {Proc.
  Natl. Acad. Sci. U.S.A.}\ }\textbf {\bibinfo {volume} {106}},\ \bibinfo
  {pages} {15320} (\bibinfo {year} {2009})}\BibitemShut {NoStop}%
\bibitem [{\citenamefont {Turlier}\ \emph {et~al.}(2016)\citenamefont
  {Turlier}, \citenamefont {Fedosov}, \citenamefont {Audoly}, \citenamefont
  {Auth}, \citenamefont {Gov}, \citenamefont {Sylkes}, \citenamefont {Joanny},
  \citenamefont {Gompper},\ and\ \citenamefont {Betz}}]{Turlier2016}%
  \BibitemOpen
  \bibfield  {author} {\bibinfo {author} {\bibfnamefont {H.}~\bibnamefont
  {Turlier}}, \bibinfo {author} {\bibfnamefont {D.~A.}\ \bibnamefont
  {Fedosov}}, \bibinfo {author} {\bibfnamefont {B.}~\bibnamefont {Audoly}},
  \bibinfo {author} {\bibfnamefont {T.}~\bibnamefont {Auth}}, \bibinfo {author}
  {\bibfnamefont {N.~S.}\ \bibnamefont {Gov}}, \bibinfo {author} {\bibfnamefont
  {C.}~\bibnamefont {Sylkes}}, \bibinfo {author} {\bibfnamefont {J.-F.}\
  \bibnamefont {Joanny}}, \bibinfo {author} {\bibfnamefont {G.}~\bibnamefont
  {Gompper}}, \ and\ \bibinfo {author} {\bibfnamefont {T.}~\bibnamefont
  {Betz}},\ }\href {\doibase 10.1038/nphys3621} {\bibfield  {journal} {\bibinfo
   {journal} {Nat. Phys.}\ }\textbf {\bibinfo {volume} {12}},\ \bibinfo {pages}
  {513} (\bibinfo {year} {2016})}\BibitemShut {NoStop}%
\bibitem [{\citenamefont {Ben-Isaac}\ \emph {et~al.}(2011)\citenamefont
  {Ben-Isaac}, \citenamefont {Park}, \citenamefont {Popescu}, \citenamefont
  {Brown}, \citenamefont {Gov},\ and\ \citenamefont {Shokef}}]{Ben-Isaac2011}%
  \BibitemOpen
  \bibfield  {author} {\bibinfo {author} {\bibfnamefont {E.}~\bibnamefont
  {Ben-Isaac}}, \bibinfo {author} {\bibfnamefont {Y.}~\bibnamefont {Park}},
  \bibinfo {author} {\bibfnamefont {G.}~\bibnamefont {Popescu}}, \bibinfo
  {author} {\bibfnamefont {F.~L.}\ \bibnamefont {Brown}}, \bibinfo {author}
  {\bibfnamefont {N.~S.}\ \bibnamefont {Gov}}, \ and\ \bibinfo {author}
  {\bibfnamefont {Y.}~\bibnamefont {Shokef}},\ }\href {\doibase
  10.1103/PhysRevLett.106.238103} {\bibfield  {journal} {\bibinfo  {journal}
  {Phys. Rev. Lett.}\ }\textbf {\bibinfo {volume} {106}},\ \bibinfo {pages} {1}
  (\bibinfo {year} {2011})},\ \Eprint {http://arxiv.org/abs/1102.4508}
  {arXiv:1102.4508} \BibitemShut {NoStop}%
\bibitem [{\citenamefont {Needleman}\ and\ \citenamefont
  {Dogic}(2017)}]{Needleman2017b}%
  \BibitemOpen
  \bibfield  {author} {\bibinfo {author} {\bibfnamefont {D.}~\bibnamefont
  {Needleman}}\ and\ \bibinfo {author} {\bibfnamefont {Z.}~\bibnamefont
  {Dogic}},\ }\href {\doibase 10.1038/natrevmats.2017.48} {\bibfield  {journal}
  {\bibinfo  {journal} {Nat. Rev. Mater.}\ }\textbf {\bibinfo {volume} {2}}
  (\bibinfo {year} {2017}),\ 10.1038/natrevmats.2017.48}\BibitemShut {NoStop}%
\bibitem [{\citenamefont {Suzuki}\ \emph {et~al.}(2015)\citenamefont {Suzuki},
  \citenamefont {Weber}, \citenamefont {Frey},\ and\ \citenamefont
  {Bausch}}]{Suzuki2015}%
  \BibitemOpen
  \bibfield  {author} {\bibinfo {author} {\bibfnamefont {R.}~\bibnamefont
  {Suzuki}}, \bibinfo {author} {\bibfnamefont {C.~A.}\ \bibnamefont {Weber}},
  \bibinfo {author} {\bibfnamefont {E.}~\bibnamefont {Frey}}, \ and\ \bibinfo
  {author} {\bibfnamefont {A.~R.}\ \bibnamefont {Bausch}},\ }\href {\doibase
  10.1080/1478422X.2017.1393245} {\bibfield  {journal} {\bibinfo  {journal}
  {Nat. Phys.}\ }\textbf {\bibinfo {volume} {11}},\ \bibinfo {pages} {839}
  (\bibinfo {year} {2015})}\BibitemShut {NoStop}%
\bibitem [{\citenamefont {Battle}\ \emph {et~al.}(2016)\citenamefont {Battle},
  \citenamefont {Broedersz}, \citenamefont {Fakhri}, \citenamefont {Geyer},
  \citenamefont {Howard}, \citenamefont {Schmidt},\ and\ \citenamefont
  {MacKintosh}}]{Battle2016}%
  \BibitemOpen
  \bibfield  {author} {\bibinfo {author} {\bibfnamefont {C.}~\bibnamefont
  {Battle}}, \bibinfo {author} {\bibfnamefont {C.~P.}\ \bibnamefont
  {Broedersz}}, \bibinfo {author} {\bibfnamefont {N.}~\bibnamefont {Fakhri}},
  \bibinfo {author} {\bibfnamefont {V.~F.}\ \bibnamefont {Geyer}}, \bibinfo
  {author} {\bibfnamefont {J.}~\bibnamefont {Howard}}, \bibinfo {author}
  {\bibfnamefont {C.~F.}\ \bibnamefont {Schmidt}}, \ and\ \bibinfo {author}
  {\bibfnamefont {F.~C.}\ \bibnamefont {MacKintosh}},\ }\href {\doibase
  10.1126/science.aac8167} {\bibfield  {journal} {\bibinfo  {journal}
  {Science.}\ }\textbf {\bibinfo {volume} {352}},\ \bibinfo {pages} {604}
  (\bibinfo {year} {2016})}\BibitemShut {NoStop}%
\bibitem [{\citenamefont {Gladrow}\ \emph {et~al.}(2016)\citenamefont
  {Gladrow}, \citenamefont {Fakhri}, \citenamefont {MacKintosh}, \citenamefont
  {Schmidt},\ and\ \citenamefont {Broedersz}}]{Gladrow2016}%
  \BibitemOpen
  \bibfield  {author} {\bibinfo {author} {\bibfnamefont {J.}~\bibnamefont
  {Gladrow}}, \bibinfo {author} {\bibfnamefont {N.}~\bibnamefont {Fakhri}},
  \bibinfo {author} {\bibfnamefont {F.~C.}\ \bibnamefont {MacKintosh}},
  \bibinfo {author} {\bibfnamefont {C.~F.}\ \bibnamefont {Schmidt}}, \ and\
  \bibinfo {author} {\bibfnamefont {C.~P.}\ \bibnamefont {Broedersz}},\ }\href
  {\doibase 10.1103/PhysRevLett.116.248301} {\bibfield  {journal} {\bibinfo
  {journal} {Phys. Rev. Lett.}\ }\textbf {\bibinfo {volume} {116}},\ \bibinfo
  {pages} {248301} (\bibinfo {year} {2016})}\BibitemShut {NoStop}%
\bibitem [{\citenamefont {Gladrow}\ \emph {et~al.}(2017)\citenamefont
  {Gladrow}, \citenamefont {Broedersz},\ and\ \citenamefont
  {Schmidt}}]{Gladrow2017}%
  \BibitemOpen
  \bibfield  {author} {\bibinfo {author} {\bibfnamefont {J.}~\bibnamefont
  {Gladrow}}, \bibinfo {author} {\bibfnamefont {C.~P.}\ \bibnamefont
  {Broedersz}}, \ and\ \bibinfo {author} {\bibfnamefont {C.~F.}\ \bibnamefont
  {Schmidt}},\ }\href {\doibase 10.1103/PhysRevE.96.022408} {\bibfield
  {journal} {\bibinfo  {journal} {Phys. Rev. E}\ }\textbf {\bibinfo {volume}
  {96}},\ \bibinfo {pages} {022408} (\bibinfo {year} {2017})},\ \Eprint
  {http://arxiv.org/abs/1704.06243} {arXiv:1704.06243} \BibitemShut {NoStop}%
\bibitem [{\citenamefont {Mura}\ \emph {et~al.}(2018)\citenamefont {Mura},
  \citenamefont {Gradziuk},\ and\ \citenamefont {Broedersz}}]{Mura2018}%
  \BibitemOpen
  \bibfield  {author} {\bibinfo {author} {\bibfnamefont {F.}~\bibnamefont
  {Mura}}, \bibinfo {author} {\bibfnamefont {G.}~\bibnamefont {Gradziuk}}, \
  and\ \bibinfo {author} {\bibfnamefont {C.~P.}\ \bibnamefont {Broedersz}},\
  }\href {\doibase 10.1103/PhysRevLett.121.038002} {\bibfield  {journal}
  {\bibinfo  {journal} {Phys. Rev. Lett.}\ }\textbf {\bibinfo {volume} {121}},\
  \bibinfo {pages} {38002} (\bibinfo {year} {2018})},\ \Eprint
  {http://arxiv.org/abs/1803.02797} {arXiv:1803.02797} \BibitemShut {NoStop}%
\bibitem [{\citenamefont {Gnesotto}\ \emph
  {et~al.}(2018{\natexlab{b}})\citenamefont {Gnesotto}, \citenamefont
  {Remlein},\ and\ \citenamefont {Broedersz}}]{Gnesotto2018}%
  \BibitemOpen
  \bibfield  {author} {\bibinfo {author} {\bibfnamefont {F.~S.}\ \bibnamefont
  {Gnesotto}}, \bibinfo {author} {\bibfnamefont {B.~M.}\ \bibnamefont
  {Remlein}}, \ and\ \bibinfo {author} {\bibfnamefont {C.~P.}\ \bibnamefont
  {Broedersz}},\ }\href {http://arxiv.org/abs/1809.04639} {\  (\bibinfo {year}
  {2018}{\natexlab{b}})},\ \Eprint {http://arxiv.org/abs/1809.04639v1}
  {arXiv:1809.04639v1} \BibitemShut {NoStop}%
\bibitem [{\citenamefont {Li}\ \emph {et~al.}(2018)\citenamefont {Li},
  \citenamefont {Horowitz}, \citenamefont {Gingrich},\ and\ \citenamefont
  {Fakhri}}]{Li2018}%
  \BibitemOpen
  \bibfield  {author} {\bibinfo {author} {\bibfnamefont {J.}~\bibnamefont
  {Li}}, \bibinfo {author} {\bibfnamefont {J.~M.}\ \bibnamefont {Horowitz}},
  \bibinfo {author} {\bibfnamefont {T.~R.}\ \bibnamefont {Gingrich}}, \ and\
  \bibinfo {author} {\bibfnamefont {N.}~\bibnamefont {Fakhri}},\ }\href
  {http://arxiv.org/abs/1809.02118} {\ ,\ \bibinfo {pages} {1} (\bibinfo {year}
  {2018})},\ \Eprint {http://arxiv.org/abs/1809.02118} {arXiv:1809.02118}
  \BibitemShut {NoStop}%
\bibitem [{\citenamefont {Mizuno}\ \emph {et~al.}(2007)\citenamefont {Mizuno},
  \citenamefont {Tardin}, \citenamefont {Schmidt},\ and\ \citenamefont
  {MacKintosh}}]{Mizuno2007}%
  \BibitemOpen
  \bibfield  {author} {\bibinfo {author} {\bibfnamefont {D.}~\bibnamefont
  {Mizuno}}, \bibinfo {author} {\bibfnamefont {C.}~\bibnamefont {Tardin}},
  \bibinfo {author} {\bibfnamefont {C.~F.}\ \bibnamefont {Schmidt}}, \ and\
  \bibinfo {author} {\bibfnamefont {F.~C.}\ \bibnamefont {MacKintosh}},\ }\href
  {\doibase 10.1126/science.1134404} {\bibfield  {journal} {\bibinfo  {journal}
  {Science.}\ }\textbf {\bibinfo {volume} {315}},\ \bibinfo {pages} {370}
  (\bibinfo {year} {2007})}\BibitemShut {NoStop}%
\bibitem [{\citenamefont {Martin}\ \emph {et~al.}(2001)\citenamefont {Martin},
  \citenamefont {Hudspeth},\ and\ \citenamefont {J{\"{u}}licher}}]{Martin2001}%
  \BibitemOpen
  \bibfield  {author} {\bibinfo {author} {\bibfnamefont {P.}~\bibnamefont
  {Martin}}, \bibinfo {author} {\bibfnamefont {A.~J.}\ \bibnamefont
  {Hudspeth}}, \ and\ \bibinfo {author} {\bibfnamefont {F.}~\bibnamefont
  {J{\"{u}}licher}},\ }\href {\doibase 10.1073/pnas.251530598} {\bibfield
  {journal} {\bibinfo  {journal} {Proc. Natl. Acad. Sci. U.S.A.}\ }\textbf
  {\bibinfo {volume} {98}},\ \bibinfo {pages} {14380} (\bibinfo {year}
  {2001})}\BibitemShut {NoStop}%
\bibitem [{\citenamefont {Fodor}\ \emph {et~al.}(2015)\citenamefont {Fodor},
  \citenamefont {Guo}, \citenamefont {Gov}, \citenamefont {Visco},
  \citenamefont {Weitz},\ and\ \citenamefont {van Wijland}}]{Fodor2015}%
  \BibitemOpen
  \bibfield  {author} {\bibinfo {author} {\bibfnamefont {{\'{E}}.}~\bibnamefont
  {Fodor}}, \bibinfo {author} {\bibfnamefont {M.}~\bibnamefont {Guo}}, \bibinfo
  {author} {\bibfnamefont {N.~S.}\ \bibnamefont {Gov}}, \bibinfo {author}
  {\bibfnamefont {P.}~\bibnamefont {Visco}}, \bibinfo {author} {\bibfnamefont
  {D.~A.}\ \bibnamefont {Weitz}}, \ and\ \bibinfo {author} {\bibfnamefont
  {F.}~\bibnamefont {van Wijland}},\ }\href@noop {} {\bibfield  {journal}
  {\bibinfo  {journal} {EPL (Europhysics Lett.}\ }\textbf {\bibinfo {volume}
  {110}},\ \bibinfo {pages} {48005} (\bibinfo {year} {2015})}\BibitemShut
  {NoStop}%
\bibitem [{\citenamefont {{Job Boekhoven, Wouter E. Hendriksen, Ger J. M.
  Koper, Rienk Eelkema}}(2015)}]{Boekhoven2015}%
  \BibitemOpen
  \bibfield  {author} {\bibinfo {author} {\bibfnamefont {J.~H. v.~E.}\
  \bibnamefont {{Job Boekhoven, Wouter E. Hendriksen, Ger J. M. Koper, Rienk
  Eelkema}}},\ }\href@noop {} {\bibfield  {journal} {\bibinfo  {journal}
  {Science.}\ }\textbf {\bibinfo {volume} {349}},\ \bibinfo {pages} {1075}
  (\bibinfo {year} {2015})}\BibitemShut {NoStop}%
\bibitem [{\citenamefont {Palacci}\ \emph {et~al.}(2013)\citenamefont
  {Palacci}, \citenamefont {Sacanna}, \citenamefont {Steinberg}, \citenamefont
  {Pine},\ and\ \citenamefont {Chaikin}}]{Palacci2013}%
  \BibitemOpen
  \bibfield  {author} {\bibinfo {author} {\bibfnamefont {J.}~\bibnamefont
  {Palacci}}, \bibinfo {author} {\bibfnamefont {S.}~\bibnamefont {Sacanna}},
  \bibinfo {author} {\bibfnamefont {A.~P.}\ \bibnamefont {Steinberg}}, \bibinfo
  {author} {\bibfnamefont {D.~J.}\ \bibnamefont {Pine}}, \ and\ \bibinfo
  {author} {\bibfnamefont {P.~M.}\ \bibnamefont {Chaikin}},\ }\href {\doibase
  10.1126/science.1230020} {\bibfield  {journal} {\bibinfo  {journal}
  {Science.}\ }\textbf {\bibinfo {volume} {339}},\ \bibinfo {pages} {936}
  (\bibinfo {year} {2013})}\BibitemShut {NoStop}%
\bibitem [{\citenamefont {Liu}\ \emph {et~al.}(2006)\citenamefont {Liu},
  \citenamefont {Gardel}, \citenamefont {Kroy}, \citenamefont {Frey},
  \citenamefont {Hoffman}, \citenamefont {Crocker}, \citenamefont {Bausch},\
  and\ \citenamefont {Weitz}}]{Liu2006}%
  \BibitemOpen
  \bibfield  {author} {\bibinfo {author} {\bibfnamefont {J.}~\bibnamefont
  {Liu}}, \bibinfo {author} {\bibfnamefont {M.~L.}\ \bibnamefont {Gardel}},
  \bibinfo {author} {\bibfnamefont {K.}~\bibnamefont {Kroy}}, \bibinfo {author}
  {\bibfnamefont {E.}~\bibnamefont {Frey}}, \bibinfo {author} {\bibfnamefont
  {B.~D.}\ \bibnamefont {Hoffman}}, \bibinfo {author} {\bibfnamefont {J.~C.}\
  \bibnamefont {Crocker}}, \bibinfo {author} {\bibfnamefont {A.~R.}\
  \bibnamefont {Bausch}}, \ and\ \bibinfo {author} {\bibfnamefont {D.~A.}\
  \bibnamefont {Weitz}},\ }\href {\doibase 10.1103/PhysRevLett.96.118104}
  {\bibfield  {journal} {\bibinfo  {journal} {Phys. Rev. Lett.}\ }\textbf
  {\bibinfo {volume} {96}},\ \bibinfo {pages} {1} (\bibinfo {year}
  {2006})}\BibitemShut {NoStop}%
\bibitem [{\citenamefont {Koenderink}\ \emph {et~al.}(2009)\citenamefont
  {Koenderink}, \citenamefont {Dogic}, \citenamefont {Nakamura}, \citenamefont
  {Bendix}, \citenamefont {MacKintosh}, \citenamefont {Hartwig}, \citenamefont
  {Stossel},\ and\ \citenamefont {Weitz}}]{Koenderink2009}%
  \BibitemOpen
  \bibfield  {author} {\bibinfo {author} {\bibfnamefont {G.~H.}\ \bibnamefont
  {Koenderink}}, \bibinfo {author} {\bibfnamefont {Z.}~\bibnamefont {Dogic}},
  \bibinfo {author} {\bibfnamefont {F.}~\bibnamefont {Nakamura}}, \bibinfo
  {author} {\bibfnamefont {P.~M.}\ \bibnamefont {Bendix}}, \bibinfo {author}
  {\bibfnamefont {F.~C.}\ \bibnamefont {MacKintosh}}, \bibinfo {author}
  {\bibfnamefont {J.~H.}\ \bibnamefont {Hartwig}}, \bibinfo {author}
  {\bibfnamefont {T.~P.}\ \bibnamefont {Stossel}}, \ and\ \bibinfo {author}
  {\bibfnamefont {D.~A.}\ \bibnamefont {Weitz}},\ }\href {\doibase
  10.1073/pnas.0903974106} {\bibfield  {journal} {\bibinfo  {journal} {Proc.
  Natl. Acad. Sci. U.S.A.}\ }\textbf {\bibinfo {volume} {106}},\ \bibinfo
  {pages} {15192} (\bibinfo {year} {2009})}\BibitemShut {NoStop}%
\bibitem [{\citenamefont {Alvarado}\ \emph {et~al.}(2017)\citenamefont
  {Alvarado}, \citenamefont {Sheinman}, \citenamefont {Sharma}, \citenamefont
  {MacKintosh},\ and\ \citenamefont {Koenderink}}]{Alvarado2017}%
  \BibitemOpen
  \bibfield  {author} {\bibinfo {author} {\bibfnamefont {J.}~\bibnamefont
  {Alvarado}}, \bibinfo {author} {\bibfnamefont {M.}~\bibnamefont {Sheinman}},
  \bibinfo {author} {\bibfnamefont {A.}~\bibnamefont {Sharma}}, \bibinfo
  {author} {\bibfnamefont {F.~C.}\ \bibnamefont {MacKintosh}}, \ and\ \bibinfo
  {author} {\bibfnamefont {G.~H.}\ \bibnamefont {Koenderink}},\ }\href
  {\doibase 10.1039/c7sm00834a} {\bibfield  {journal} {\bibinfo  {journal}
  {Soft Matter}\ }\textbf {\bibinfo {volume} {13}},\ \bibinfo {pages} {5624}
  (\bibinfo {year} {2017})},\ \Eprint {http://arxiv.org/abs/1612.08713}
  {arXiv:1612.08713} \BibitemShut {NoStop}%
\bibitem [{\citenamefont {Bertrand}\ \emph {et~al.}(2012)\citenamefont
  {Bertrand}, \citenamefont {Fygenson},\ and\ \citenamefont
  {Saleh}}]{Bertrand2012}%
  \BibitemOpen
  \bibfield  {author} {\bibinfo {author} {\bibfnamefont {O.~J.~N.}\
  \bibnamefont {Bertrand}}, \bibinfo {author} {\bibfnamefont {D.~K.}\
  \bibnamefont {Fygenson}}, \ and\ \bibinfo {author} {\bibfnamefont {O.~A.}\
  \bibnamefont {Saleh}},\ }\href {\doibase 10.1073/pnas.1208732109} {\bibfield
  {journal} {\bibinfo  {journal} {Proc. Natl. Acad. Sci. U.S.A.}\ }\textbf
  {\bibinfo {volume} {109}},\ \bibinfo {pages} {17342} (\bibinfo {year}
  {2012})}\BibitemShut {NoStop}%
\bibitem [{\citenamefont {Seifert}(2012)}]{Seifertreview}%
  \BibitemOpen
  \bibfield  {author} {\bibinfo {author} {\bibfnamefont {U.}~\bibnamefont
  {Seifert}},\ }\href {\doibase 10.1088/0034-4885/75/12/126001} {\bibfield
  {journal} {\bibinfo  {journal} {Reports Prog. Phys.}\ }\textbf {\bibinfo
  {volume} {75}},\ \bibinfo {pages} {126001} (\bibinfo {year}
  {2012})}\BibitemShut {NoStop}%
\bibitem [{\citenamefont {Frishman}\ and\ \citenamefont
  {Ronceray}(2018)}]{Frishman2018}%
  \BibitemOpen
  \bibfield  {author} {\bibinfo {author} {\bibfnamefont {A.}~\bibnamefont
  {Frishman}}\ and\ \bibinfo {author} {\bibfnamefont {P.}~\bibnamefont
  {Ronceray}},\ }\href@noop {} {\  (\bibinfo {year} {2018})},\ \Eprint
  {http://arxiv.org/abs/1809.09650} {arXiv:1809.09650} \BibitemShut {NoStop}%
\bibitem [{\citenamefont {Seara}\ \emph {et~al.}(2018)\citenamefont {Seara},
  \citenamefont {Yadav}, \citenamefont {Linsmeier}, \citenamefont {Tabatabai},
  \citenamefont {Oakes}, \citenamefont {Tabei}, \citenamefont {Banerjee},\ and\
  \citenamefont {Murrell}}]{Seara2018}%
  \BibitemOpen
  \bibfield  {author} {\bibinfo {author} {\bibfnamefont {D.~S.}\ \bibnamefont
  {Seara}}, \bibinfo {author} {\bibfnamefont {V.}~\bibnamefont {Yadav}},
  \bibinfo {author} {\bibfnamefont {I.}~\bibnamefont {Linsmeier}}, \bibinfo
  {author} {\bibfnamefont {A.~P.}\ \bibnamefont {Tabatabai}}, \bibinfo {author}
  {\bibfnamefont {P.~W.}\ \bibnamefont {Oakes}}, \bibinfo {author}
  {\bibfnamefont {S.~M.~A.}\ \bibnamefont {Tabei}}, \bibinfo {author}
  {\bibfnamefont {S.}~\bibnamefont {Banerjee}}, \ and\ \bibinfo {author}
  {\bibfnamefont {M.~P.}\ \bibnamefont {Murrell}},\ }\href {\doibase
  10.1038/s41467-018-07413-5} {\bibfield  {journal} {\bibinfo  {journal} {Nat.
  Commun.}\ }\textbf {\bibinfo {volume} {9}},\ \bibinfo {pages} {4948}
  (\bibinfo {year} {2018})},\ \Eprint {http://arxiv.org/abs/1804.04232}
  {arXiv:1804.04232} \BibitemShut {NoStop}%
\bibitem [{\citenamefont {Rold{\'{a}}n}\ \emph {et~al.}(2018)\citenamefont
  {Rold{\'{a}}n}, \citenamefont {Barral}, \citenamefont {Martin}, \citenamefont
  {Parrondo},\ and\ \citenamefont {J{\"{u}}licher}}]{Roldan2018}%
  \BibitemOpen
  \bibfield  {author} {\bibinfo {author} {\bibfnamefont {{\'{E}}.}~\bibnamefont
  {Rold{\'{a}}n}}, \bibinfo {author} {\bibfnamefont {J.}~\bibnamefont
  {Barral}}, \bibinfo {author} {\bibfnamefont {P.}~\bibnamefont {Martin}},
  \bibinfo {author} {\bibfnamefont {J.~M.~R.}\ \bibnamefont {Parrondo}}, \ and\
  \bibinfo {author} {\bibfnamefont {F.}~\bibnamefont {J{\"{u}}licher}},\ }\href
  {\doibase arXiv:1803.04743v1} {\ \textbf {\bibinfo {volume} {2}},\ \bibinfo
  {pages} {1} (\bibinfo {year} {2018})},\ \Eprint
  {http://arxiv.org/abs/1803.04743} {arXiv:1803.04743} \BibitemShut {NoStop}%
\bibitem [{\citenamefont {Bisker}\ \emph {et~al.}(2017)\citenamefont {Bisker},
  \citenamefont {Polettini}, \citenamefont {Gingrich},\ and\ \citenamefont
  {Horowitz}}]{Bisker2017}%
  \BibitemOpen
  \bibfield  {author} {\bibinfo {author} {\bibfnamefont {G.}~\bibnamefont
  {Bisker}}, \bibinfo {author} {\bibfnamefont {M.}~\bibnamefont {Polettini}},
  \bibinfo {author} {\bibfnamefont {T.~R.}\ \bibnamefont {Gingrich}}, \ and\
  \bibinfo {author} {\bibfnamefont {J.~M.}\ \bibnamefont {Horowitz}},\
  }\href@noop {} {\bibfield  {journal} {\bibinfo  {journal} {J. Stat. Mech.
  Theory Exp.}\ }\textbf {\bibinfo {volume} {2017}} (\bibinfo {year} {2017})},\
  \Eprint {http://arxiv.org/abs/1708.06769} {arXiv:1708.06769} \BibitemShut
  {NoStop}%
\bibitem [{\citenamefont {Esposito}(2012)}]{Esposito2012}%
  \BibitemOpen
  \bibfield  {author} {\bibinfo {author} {\bibfnamefont {M.}~\bibnamefont
  {Esposito}},\ }\href@noop {} {\bibfield  {journal} {\bibinfo  {journal}
  {Phys. Rev. E}\ }\textbf {\bibinfo {volume} {85}},\ \bibinfo {pages} {41125}
  (\bibinfo {year} {2012})}\BibitemShut {NoStop}%
\bibitem [{\citenamefont {Polettini}\ and\ \citenamefont
  {Esposito}(2017)}]{Polettini2017}%
  \BibitemOpen
  \bibfield  {author} {\bibinfo {author} {\bibfnamefont {M.}~\bibnamefont
  {Polettini}}\ and\ \bibinfo {author} {\bibfnamefont {M.}~\bibnamefont
  {Esposito}},\ }\href {\doibase 10.1103/PhysRevLett.119.240601} {\bibfield
  {journal} {\bibinfo  {journal} {Phys. Rev. Lett.}\ }\textbf {\bibinfo
  {volume} {119}},\ \bibinfo {pages} {240601} (\bibinfo {year}
  {2017})}\BibitemShut {NoStop}%
\bibitem [{\citenamefont {Ghanta}\ \emph {et~al.}(2017)\citenamefont {Ghanta},
  \citenamefont {Neu},\ and\ \citenamefont {Teitsworth}}]{Ghanta2017}%
  \BibitemOpen
  \bibfield  {author} {\bibinfo {author} {\bibfnamefont {A.}~\bibnamefont
  {Ghanta}}, \bibinfo {author} {\bibfnamefont {J.~C.}\ \bibnamefont {Neu}}, \
  and\ \bibinfo {author} {\bibfnamefont {S.}~\bibnamefont {Teitsworth}},\
  }\href {\doibase 10.1103/PhysRevE.95.032128} {\bibfield  {journal} {\bibinfo
  {journal} {Phys. Rev. E}\ }\textbf {\bibinfo {volume} {95}},\ \bibinfo
  {pages} {1} (\bibinfo {year} {2017})}\BibitemShut {NoStop}%
\bibitem [{\citenamefont {Gonzalez}\ \emph {et~al.}(2018)\citenamefont
  {Gonzalez}, \citenamefont {Neu},\ and\ \citenamefont
  {Teitsworth}}]{Gonzalez2018}%
  \BibitemOpen
  \bibfield  {author} {\bibinfo {author} {\bibfnamefont {J.~P.}\ \bibnamefont
  {Gonzalez}}, \bibinfo {author} {\bibfnamefont {J.~C.}\ \bibnamefont {Neu}}, \
  and\ \bibinfo {author} {\bibfnamefont {S.~W.}\ \bibnamefont {Teitsworth}},\
  }\href@noop {} {\ ,\ \bibinfo {pages} {1} (\bibinfo {year} {2018})},\ \Eprint
  {http://arxiv.org/abs/1810.07865v1} {arXiv:1810.07865v1} \BibitemShut
  {NoStop}%
\bibitem [{\citenamefont {Yucht}\ \emph {et~al.}(2013)\citenamefont {Yucht},
  \citenamefont {Sheinman},\ and\ \citenamefont {Broedersz}}]{Yucht2013}%
  \BibitemOpen
  \bibfield  {author} {\bibinfo {author} {\bibfnamefont {M.~G.}\ \bibnamefont
  {Yucht}}, \bibinfo {author} {\bibfnamefont {M.}~\bibnamefont {Sheinman}}, \
  and\ \bibinfo {author} {\bibfnamefont {C.~P.}\ \bibnamefont {Broedersz}},\
  }\href {\doibase 10.1039/c3sm50177a} {\bibfield  {journal} {\bibinfo
  {journal} {Soft Matter}\ }\textbf {\bibinfo {volume} {9}},\ \bibinfo {pages}
  {7000} (\bibinfo {year} {2013})},\ \Eprint {http://arxiv.org/abs/1301.3574}
  {arXiv:1301.3574} \BibitemShut {NoStop}%
\bibitem [{\citenamefont {Broedersz}\ and\ \citenamefont
  {Mackintosh}(2014)}]{Broedersz2014}%
  \BibitemOpen
  \bibfield  {author} {\bibinfo {author} {\bibfnamefont {C.~P.}\ \bibnamefont
  {Broedersz}}\ and\ \bibinfo {author} {\bibfnamefont {F.~C.}\ \bibnamefont
  {Mackintosh}},\ }\href {\doibase 10.1103/RevModPhys.86.995} {\bibfield
  {journal} {\bibinfo  {journal} {Rev. Mod. Phys.}\ }\textbf {\bibinfo {volume}
  {86}},\ \bibinfo {pages} {995} (\bibinfo {year} {2014})},\ \Eprint
  {http://arxiv.org/abs/1404.4332} {arXiv:1404.4332} \BibitemShut {NoStop}%
\bibitem [{\citenamefont {Osmanovi{\'{c}}}\ and\ \citenamefont
  {Rabin}(2017)}]{Osmanovic2017}%
  \BibitemOpen
  \bibfield  {author} {\bibinfo {author} {\bibfnamefont {D.}~\bibnamefont
  {Osmanovi{\'{c}}}}\ and\ \bibinfo {author} {\bibfnamefont {Y.}~\bibnamefont
  {Rabin}},\ }\href {\doibase 10.1039/C6SM02722A} {\bibfield  {journal}
  {\bibinfo  {journal} {Soft Matter}\ }\textbf {\bibinfo {volume} {13}},\
  \bibinfo {pages} {963} (\bibinfo {year} {2017})},\ \Eprint
  {http://arxiv.org/abs/1608.05914} {arXiv:1608.05914} \BibitemShut {NoStop}%
\bibitem [{\citenamefont {Mao}\ and\ \citenamefont {Lubensky}(2018)}]{Mao2018}%
  \BibitemOpen
  \bibfield  {author} {\bibinfo {author} {\bibfnamefont {X.}~\bibnamefont
  {Mao}}\ and\ \bibinfo {author} {\bibfnamefont {T.~C.}\ \bibnamefont
  {Lubensky}},\ }\href {\doibase 10.1146/annurev-conmatphys-033117-054235}
  {\bibfield  {journal} {\bibinfo  {journal} {Annu. Rev. Condens. Matter
  Phys.}\ }\textbf {\bibinfo {volume} {9}},\ \bibinfo {pages} {413} (\bibinfo
  {year} {2018})}\BibitemShut {NoStop}%
\bibitem [{\citenamefont {Falasco}\ \emph {et~al.}(2015)\citenamefont
  {Falasco}, \citenamefont {Baiesi}, \citenamefont {Molinaro}, \citenamefont
  {Conti},\ and\ \citenamefont {Baldovin}}]{Falasco2015}%
  \BibitemOpen
  \bibfield  {author} {\bibinfo {author} {\bibfnamefont {G.}~\bibnamefont
  {Falasco}}, \bibinfo {author} {\bibfnamefont {M.}~\bibnamefont {Baiesi}},
  \bibinfo {author} {\bibfnamefont {L.}~\bibnamefont {Molinaro}}, \bibinfo
  {author} {\bibfnamefont {L.}~\bibnamefont {Conti}}, \ and\ \bibinfo {author}
  {\bibfnamefont {F.}~\bibnamefont {Baldovin}},\ }\href {\doibase
  10.1103/PhysRevE.92.022129} {\bibfield  {journal} {\bibinfo  {journal} {Phys.
  Rev. E - Stat. Nonlinear, Soft Matter Phys.}\ }\textbf {\bibinfo {volume}
  {92}},\ \bibinfo {pages} {1} (\bibinfo {year} {2015})},\ \Eprint
  {http://arxiv.org/abs/1505.05088v1} {arXiv:1505.05088v1} \BibitemShut
  {NoStop}%
\bibitem [{\citenamefont {Risken}(1996)}]{Risken}%
  \BibitemOpen
  \bibfield  {author} {\bibinfo {author} {\bibfnamefont {H.}~\bibnamefont
  {Risken}},\ }\href@noop {} {\emph {\bibinfo {title} {{The Fokker-Planck
  equation}}}},\ edited by\ \bibinfo {editor} {\bibnamefont {Springer}}\
  (\bibinfo {address} {Berlin},\ \bibinfo {year} {1996})\BibitemShut {NoStop}%
\bibitem [{\citenamefont {Weiss}(2003)}]{Weiss2003}%
  \BibitemOpen
  \bibfield  {author} {\bibinfo {author} {\bibfnamefont {J.~B.}\ \bibnamefont
  {Weiss}},\ }\href {\doibase 10.1034/j.1600-0870.2003.00014.x} {\bibfield
  {journal} {\bibinfo  {journal} {Tellus A}\ }\textbf {\bibinfo {volume}
  {55}},\ \bibinfo {pages} {208} (\bibinfo {year} {2003})}\BibitemShut
  {NoStop}%
\bibitem [{\citenamefont {{Mohar, B and Alavi, Y and Chartrand, G and
  Oellermann, Ortrud and Schwenk}}(1991)}]{Mohar1991}%
  \BibitemOpen
  \bibfield  {author} {\bibinfo {author} {\bibfnamefont {A.}~\bibnamefont
  {{Mohar, B and Alavi, Y and Chartrand, G and Oellermann, Ortrud and
  Schwenk}}},\ }\href@noop {} {\bibfield  {journal} {\bibinfo  {journal} {Graph
  Theory, Comb. Appl.}\ }\textbf {\bibinfo {volume} {2}},\ \bibinfo {pages}
  {5364} (\bibinfo {year} {1991})}\BibitemShut {NoStop}%
\bibitem [{\citenamefont {Schaeffer}(2007)}]{Schaeffer2007}%
  \BibitemOpen
  \bibfield  {author} {\bibinfo {author} {\bibfnamefont {S.~E.}\ \bibnamefont
  {Schaeffer}},\ }\href {\doibase 10.1016/j.cosrev.2007.05.001} {\bibfield
  {journal} {\bibinfo  {journal} {Comput. Sci. Rev.}\ }\textbf {\bibinfo
  {volume} {1}},\ \bibinfo {pages} {27} (\bibinfo {year} {2007})},\ \Eprint
  {http://arxiv.org/abs/1111.7236v1} {arXiv:1111.7236v1} \BibitemShut {NoStop}%
\bibitem [{\citenamefont {Fisher}(1966)}]{Fisher1966}%
  \BibitemOpen
  \bibfield  {author} {\bibinfo {author} {\bibfnamefont {M.~E.}\ \bibnamefont
  {Fisher}},\ }\href@noop {} {\bibfield  {journal} {\bibinfo  {journal} {J.
  Comb. theory}\ }\textbf {\bibinfo {volume} {125}},\ \bibinfo {pages} {105}
  (\bibinfo {year} {1966})}\BibitemShut {NoStop}%
\bibitem [{\citenamefont {{G. Kirchhoff}}(1847)}]{Kirchhoff1847}%
  \BibitemOpen
  \bibfield  {author} {\bibinfo {author} {\bibnamefont {{G. Kirchhoff}}},\
  }\href@noop {} {\bibfield  {journal} {\bibinfo  {journal} {Ann. Phys.}\
  }\textbf {\bibinfo {volume} {12}},\ \bibinfo {pages} {497} (\bibinfo {year}
  {1847})}\BibitemShut {NoStop}%
\bibitem [{\citenamefont {Mura}\ \emph {et~al.}(2019)\citenamefont {Mura},
  \citenamefont {Gradziuk},\ and\ \citenamefont {Broedersz}}]{Mura2019}%
  \BibitemOpen
  \bibfield  {author} {\bibinfo {author} {\bibfnamefont {F.}~\bibnamefont
  {Mura}}, \bibinfo {author} {\bibfnamefont {G.}~\bibnamefont {Gradziuk}}, \
  and\ \bibinfo {author} {\bibfnamefont {C.~P.}\ \bibnamefont {Broedersz}},\
  }\href@noop {} {\bibfield  {journal} {\bibinfo  {journal} {To be Publ.}\ }
  (\bibinfo {year} {2019})}\BibitemShut {NoStop}%
\end{thebibliography}%

\onecolumngrid
\begin{appendices}
\section{Calculation for a d-dimensional zero-restlength cubic lattice}
\label{appendix1}
First we find the profile of the $\nC$ matrix for the case of a single activity at site $(0,\ldots,0)$. Neglecting the boundary conditions and assuming a rotational symmetry of the solution we find:
\begin{equation}
\nc_{\N,\Nb}=a_d\left( \sd n_i^2 +\sd \nb_i^2 \right)^{-(d-1)}\quad \mathrm{with} \quad a_d=\frac{(d-2)!}{2\pi^d}
\end{equation}
\begin{equation}
\partial_{n_i}\nc_{\N,\Nb}=-(d-1)a_d\frac{2n_i}{(n_1^2+\ldots+\nb_d^2)^d}
\end{equation}
\begin{align}
\partial_{n_i}^2\nc_{\N,\Nb}&=d(d-1)a_d\frac{4n_i^2}{(n_1^2+\ldots+\nb_d^2)^{d+1}}-(d-1)a_d\frac{2}{(n_1^2+\ldots+\nb_d^2)^{d}}=\\
&=\frac{2(d-1)a_d}{(n_1^2+\ldots+\nb_d^2)^{d+1}}\left[ 2dn_i^2-(n_1^2+\ldots+\nb_d^2) \right]
\end{align}
Adding contributions from the all the second derivatives appearing in Eq.~\eqref{eq:omegalind} we get:
\begin{equation}
\label{eq:partials}
\sd\partial_{n_i}^2\nc_{\N,\Nb}=2d(d-1)a_d\frac{[(n_1^2+\ldots+n_d^2)-(\nb_1^2+\ldots+\nb_d^2)]}{(n_1^2+\ldots+\nb_d^2)^{d+1}}=2d(d-1)a_d\frac{\N^2-\Nb^2}{(\N^2+\Nb^2)^{d+1}}
\end{equation}
In order to get an expression for $\sd\partial_{x_i}^2c$ in the case of one active bead at site $(z_1,\ldots,z_d)$, one simply has to substitute $n_i\to (n_i-z_i)$, $\nb_i\to (\nb_i-z_i)$ in Eq.~\eqref{eq:partials}.
Therefore, the contribution to $\omega$ from an activity $\alpha_{z_1,\ldots,z_d}$ at site $(z_1,\ldots,z_d)$ reads:
\begin{align}
\sd\partial_{n_i}^2c_{r,\ldots,r;-r,\ldots,-r}&=2d(d-1)a_d\frac{\sd(r-z_i)^2-\sd(-r-z_i)^2}{\left( \sd(r-z_i)^2+\sd(-r-z_i)^2 \right)^{d+1}}\alpha_{z_1,\ldots,z_d}=\\
&=2d(d-1)a_d\frac{\sd(-4rz_i)}{\left(2r^2d+2\sd z_i^2\right)^{d+1}}\alpha_{z_1,\ldots,z_d}=\\
&=-\frac{d(d-1)a_d}{2^{d-2}}\frac{\sd rz_i}{\left(r^2d+\sd z_i^2\right)^{d+1}}\alpha_{z_1,\ldots,z_d}
\end{align}
Finally, we can proceed to calculating $\mean{\omega^2(2\sqrt{d}\ r)}$.
\begin{align}
&\mean{\left( \sum_{z_1,\ldots,z_d}\frac{(\sd rz_i)\alpha_{z_1,\ldots,z_d}}{\left(r^2d+\sd z_i^2\right)^{d+1}} \right) \left( \sum_{\tz_1,\ldots,\tz_d}\frac{(\sd r\tz_i)\alpha_{\tz_1,\ldots,\tz_d}}{\left(r^2d+\sd \tz_i^2\right)^{d+1}} \right)}=
\\
&\stackrel{(1)}{=}\sum_{z_1,\ldots,z_d}\frac{(\sd rz_i)^2\sigma_\alpha^2}{\left(r^2d+\sd z_i^2\right)^{2d+2}}\stackrel{(2)}{=}\sum_{z_1,\ldots,z_d}\frac{\sd r^2z_i^2\sigma_\alpha^2}{\left(r^2d+\sd z_i^2\right)^{2d+2}}\approx\\
&\stackrel{cont.}{\approx} \int_{z_1,\ldots,z_d}\frac{\sd r^2z_i^2\sigma_\alpha^2}{\left(r^2d+\sd z_i^2\right)^{2d+2}}
\end{align}
Step $(1)$ follows from $\mean{\alpha_{z_1,\ldots,z_d} \alpha_{\tz_1,\ldots,\tz_d}}=\sigma_\alpha^2\delta_{z_1,\tz_1}\cdots\delta_{z_d,\tz_d}$. Thereby we have assumed that $	\bar{\alpha}=0$. This can be achieved by replacing the noise amplitudes $\alpha_i$ with $\alpha_i-\bar{\alpha}$. This transformation is justified, because any shift of the active noise amplitudes by a constant value does not affect $\widetilde{\partial}^2\nc_{ij}$ (compare with $\widetilde{\partial}^2 \ec_{ij}=0\ \forall_{i\neq j}$). Note that $\alpha_i-\bar{\alpha}$ are introduced just for convenience and one should not think of them as of any noise amplitudes. Step $(2)$ results from the fact that the terms odd in $z_i$ sum up to $0$. In the last step we approximated the sum by an integral.

\section{Calculation for finite restlength triangular lattice}
\label{appendix2}
\begin{figure}[H]
\centering
\scalebox{0.4}{
\includegraphics{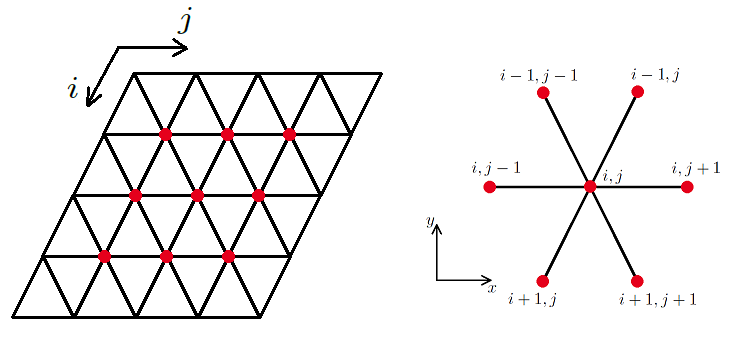}}
\caption{Triangular lattice and indexing of beads.}
\label{2Dfintri}
\end{figure}

Here we derive equation for the covariance matrix for the case of a finite restlength triangular lattice. We index the beads in the lattice as shown in Fig.~\ref{2Dfintri}
Let us denote by $f^x_{ij}$ and $f^y_{ij}$ the $x$- and $y$-components of the force acting on bead $(i,j)$, and by $x_{ij}$, $y_{ij}$ the $x$- and $y$-displacements of bead $(i,j)$. Expanding the force up to linear order in displacements we find:
\begin{align*}
f^x_{ij}=&\alpha(x_{i-1,j-1}+x_{i+1,j+1}-2x_{i,j}+x_{i-1,j}+x_{i+1,j}-2x_{i,j})+(x_{i,j-1}+x_{i,j+1}-2x_{i,j})\\
&-\beta(y_{i-1,j-1}+y_{i+1,j+1}-2y_{i,j})+\beta(y_{i-1,j}+y_{i+1,j}-2y_{i,j})\\
f^y_{ij}=&\gamma(y_{i-1,j-1}+y_{i+1,j+1}-2y_{i,j}+y_{i-1,j}+y_{i+1,j}-2y_{i,j})\\
&-\delta(x_{i-1,j-1}+x_{i+1,j+1}-2x_{i,j})+\delta(x_{i-1,j}+x_{i+1,j}-2x_{i,j})
\end{align*}
with $\alpha=1/4$, $\beta=\sqrt{3}/4$, $\gamma=3/4$, $\delta=\sqrt{3}/4$.
It is convenient to rewrite the Lyapunov equation in the following way:
\begin{equation}
-2\D=\A\C+\C\A=\mean{\A\x\x^T+\x(\A\x)^T}=\mean{\F\x^T+\x\F^T}
\end{equation}
Let us denote the elements of the covariance matrix by: $\mean{x_{ij}x_{kl}}=c^{xx}_{ij;kl}$, $\mean{x_{ij}y_{kl}}=c^{xy}_{ij;kl}$, $\mean{y_{ij}y_{kl}}=c^{yy}_{ij;kl}$ and introduce the discrete derivative operators:

\begin{equation*}
\begin{aligned}[c]
\pari c_{ij;kl}&=c_{i-1,j;k,l}-2c_{i,j;k,l}+c_{i+1,j;k,l}\\
\parj c_{ij;kl}&=c_{i,j-1;k,l}-2c_{i,j;k,l}+c_{i,j+1;k,l}\\
\pard c_{ij;kl}&=c_{i-1,j-1;k,l}-2c_{i,j;k,l}+c_{i+1,j+1;k,l}
\end{aligned}
\qquad
\begin{aligned}[c]
\parib c_{ij;kl}&=c_{i,j;k-1,l}-2c_{i,j;k,l}+c_{i,j;k+1,l}\\
\parjb c_{ij;kl}&=c_{i,j;k,l-1}-2c_{i,j;k,l}+c_{i,j;k,l+1}\\
\pardb c_{ij;kl}&=c_{i,j;k-1,l-1}-2c_{i,j;k,l}+c_{i,j;k+1,l+1}
\end{aligned}
\end{equation*}
Then, the Lyapunov equation translates to:
\begin{flalign}
\mean{f^x_{ij}x_{kl}+x_{ij}f^x_{kl}}=&[\alpha(\pard+\pari)+\parj]c^{xx}_{ij;kl}-\beta(\pard-\pari)c^{yx}_{ij;kl}&&\\\nonumber
+&[\alpha(\pardb+\parib)+\parjb]c^{xx}_{ij;kl}-\beta(\pardb-\parib)c^{xy}_{ij;kl}=-2\delta_{(ij),(kl)}d_{ij}&&
\end{flalign}
\begin{flalign}
\label{eq:syf2}
\mean{f^x_{ij}y_{kl}+x_{ij}f^y_{kl}}=&[\alpha(\pard+\pari)+\parj]c^{xy}_{ij;kl}-\beta(\pard-\pari)c^{yy}_{ij;kl}&&\\\nonumber
+&\gamma(\pardb+\parib)c^{xy}_{ij;kl}-\delta(\pardb-\parib)c^{xx}_{ij;kl}=0&&
\end{flalign}
\begin{flalign}
\label{eq:syf3}
\mean{f^x_{ij}y_{kl}+x_{ij}f^y_{kl}}=&[\alpha(\pardb+\parib)+\parjb]c^{yx}_{ij;kl}-\beta(\pardb-\parib)c^{yy}_{ij;kl}&&\\\nonumber
+&\gamma(\pard+\pari)c^{yx}_{ij;kl}-\delta(\pard-\pari)c^{xx}_{ij;kl}=0&&
\end{flalign}
\begin{flalign}
\mean{f^y_{ij}y_{kl}+y_{ij}f^y_{kl}}=&\gamma(\pard+\pari)c^{yy}_{ij;kl}-\delta(\pard-\pari)c^{xy}_{ij;kl}&&\\\nonumber
+&\gamma(\pardb+\parib)c^{yy}_{ij;kl}-\delta(\pardb-\parib)c^{yx}_{ij;kl}=-2\delta_{(ij),(kl)}d_{ij}&&
\end{flalign}
If we want to move to a continuous picture, we replace $\pari\to\partial_1^2$, $\parj\to\partial_2^2$, $\pard\to (\partial_1+\partial_2)^2$. In this picture $c^{xx}$, $c^{xy}$, $c^{yx}$, $c^{yy}$ should be seen as functions on a 4-dimensional cube. 

One can also write down equations for the cycling frequencies using:
\begin{equation}
\omega_{x_{ij},y_{kl}}=\frac{\mean{f^x_{ij}y_{kl}-x_{ij}f^y_{kl}}}{2\sqrt{\det\Cr}}\stackrel{\D-diag.}{=}\frac{\mean{f^x_{ij}y_{kl}}}{\sqrt{\det\Cr}}
\end{equation}
In the last step we used the Lyapunov equation together with the fact that $\D$ is a diagonal matrix.

\end{appendices}
\end{document}